\documentclass[sponsored]{acmsiggraph}

\usepackage{hyperref}
\usepackage{rotating}
\usepackage{xcolor}
\usepackage[abs]{overpic}
\usepackage{amsmath}
\usepackage{amssymb}

\newcommand{\boldhead}[1]{\vspace{0.05in}\noindent\textbf{#1.}}

\newcommand{\ve}[1]{{\bf #1}}
\newcommand{\comment}[1]{}
\newcommand{\urlwofont}[1] {\urlstyle{same}\url{#1}}

\TOGonlineid{}
\TOGvolume{0}
\TOGnumber{0}
\TOGarticleDOI{1111111.2222222}
\TOGprojectURL{}
\TOGvideoURL{}
\TOGdataURL{}
\TOGcodeURL{}

\title{Blind Recovery of Spatially Varying Reflectance from a Single Image}

\author{Kevin Karsch \hspace{10mm} David Forsyth\\ University of Illinois}
\pdfauthor{Kevin Karsch}


\begin{document}

\teaser{
\begin{minipage}{\textwidth}
\vspace{-3mm}
\includegraphics[width=\textwidth]{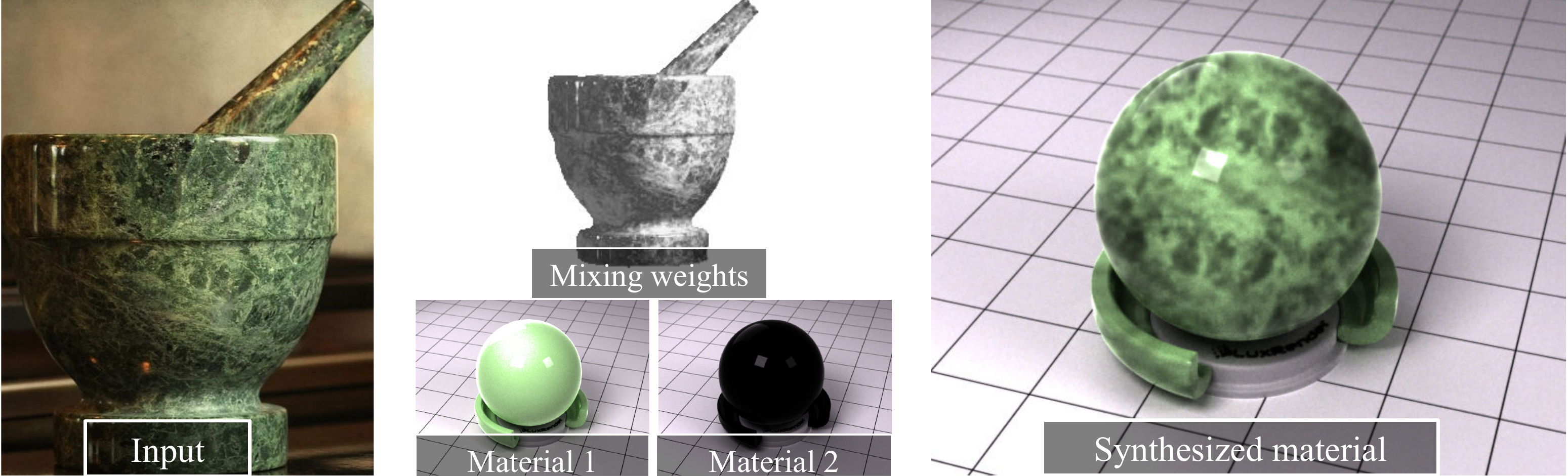}
\vspace{-6mm}
\caption{From a single photograph, our method estimates spatially varying materials (diffuse reflectance and specular parameters). The input image is decomposed into $k$ low-order, parametric materials (Material 1 and 2) and a set of per-pixel material mixing coefficients (Mixing weights); shape and illumination is jointly inferred. This decomposition can be transferred to new shapes (Synthesized material) and also used to generate new materials.
\vspace{-1mm}
}
\label{fig:teaser}
\end{minipage}
}

\maketitle

\begin{abstract}
\vspace{-3mm}
We propose a new technique for estimating spatially varying parametric materials from a single image of an object with unknown shape in unknown illumination. Our method uses a low-order parametric reflectance model, and incorporates strong assumptions about lighting and shape. We develop new priors about how materials mix over space, and jointly infer all of these properties from a single image. This produces a decomposition of an image which corresponds, in one sense, to microscopic features (material reflectance) and macroscopic features (weights defining the mixing properties of materials over space). We have built a large dataset of real objects rendered with different material models under different illumination fields for training and ground truth evaluation.  Extensive experiments on both our synthetic dataset images as well as real images show that (a) our method recovers parameters with reasonable accuracy; (b) material parameters recovered by our method give accurate predictions of new renderings of the object; and (c) our low-order reflectance model still provides a good fit to many real-world reflectances.
\end{abstract}

\begin{CRcatlist}
\CRcat{I.2.10.d}{Artificial Intelligence}{Vision and Scene Understanding}{Modeling and recovery of physical attributes};
\CRcat{I.3.8}{Computer Graphics}{Applications}{};
\CRcat{I.4.8.c}{Image Processing and Computer Vision}{Image Models}{}
\end{CRcatlist}

\keywordlist

\TOGlinkslist

\copyrightspace

\section{Introduction}
Humans are quite good at guessing an object's material based on appearance alone~\cite{Adelson99lightnessperception}. However, material\footnote{We abbreviate ``material reflectance'' with ``material.''} estimation from a single photograph remains a challenging and unsolved problem in computer vision. Appearance is often considered a function of object shape, incident illumination, and surface reflectance, and many solutions have been proposed addressing the problem of material estimation from a single image if shape and/or illumination are known precisely. 

Romeiro and Zickler first showed how to estimate reflectance under known shape and illumination~\cite{Romeiro:eccv:08}, and Romeiro et al. later extended this work by marginalizing over illumination~\cite{Romeiro:eccv:10}. Generalizing further, Lombardi and Nishino~\cite{Lombardi:eccv:2012} recover reflectance and illumination from an image assuming only that the object's shape is known, and Oxholm and Nishino~\cite{Oxholm:eccv:2012} estimate reflectance and shape under exact lighting. If multiple images are available, it is also possible to recover shape and spatially varying reflectance~\cite{Alldrin:cvpr:2008,Goldman:tpami:2010}. These techniques provide valuable intuition for moving forward, yet they hinge on knowing {\it exact} shape or {\it exact} illumination, or have strict setup requirements (directional light, multiple photos, etc), and require a fundamentally different approach when additional information is not available. 

Such approaches have been proposed by Barron and Malik~\cite{Barron:eccv:12,Barron:cvpr:12}, who use strict priors to jointly recover shape, diffuse albedo, and illumination. However, as in many shape-from-shading algorithms, all surfaces are assumed to be Lambertian. Glossy surfaces are thus impossible to recover and may cause errors in estimation. Furthermore, Lambertian models of material are not suitable for describing a large percentage of real-world surfaces, limiting the applicability of these techniques.

A major concern of prior work is in recovering real-world BRDFs and high-frequency illumination~\cite{Lombardi:eccv:2012,Oxholm:eccv:2012,Romeiro:eccv:10}, or that recovered shapes are integrable and reconstructions are exact (image and re-rendered image match exactly)~\cite{Barron:eccv:12,Barron:cvpr:12}. However, it is well known that recovering these high-parametric solutions is ill-posed in many cases, and precise conditions must be met to estimate these robustly. For example, real-world BRDFs can be extracted from a curved shape and single directional light (known a priori)~\cite{Chandraker:iccv:11}, and surface normals can be found given enough pictures with particular lighting conditions and isotropic (yet unknown) BRDFs~\cite{Alldrin:iccv:2007,Shi:eccv:2012}.

We opt for lower-order representations of reflectance and illumination. Our idea is to reduce the number of parameters we recover, relax the constraints imposed by prior methods, and attempt to recover materials from a more practical perspective. Our main goal is material inference, but we must jointly optimize over shape and illumination since these are unknown to our algorithm. We consider simple models of reflectance and lighting (models often used by artists, where perception is the only metric that matters), and impose only soft constraints on shape reconstruction. Our material model is low-order (only five parameters), allowing us to tease good estimates of materials from images, even if our recovered shape and illumination estimates are somewhat inaccurate. We also show that our model can be extended to spatially varying materials by inferring mixture coefficients (linearly combining 5-parameter materials) at each pixel. Figure~\ref{fig:teaser} demonstrates the results of our estimation technique on a marble mortar and pestle.

Most similar to our work is the method of Goldman et al. \cite{Goldman:tpami:2010}. They estimate per-pixel mixture weights and a set of parametric materials, but require multiple HDR images under {\it known} lighting and impose limiting priors on mixture weights. Our method is applicable to single, LDR images (lighting is jointly estimated), and we also develop new priors for better mixture weights. 

\begin{figure}[t]
\includegraphics[width=0.49\columnwidth,page=1]{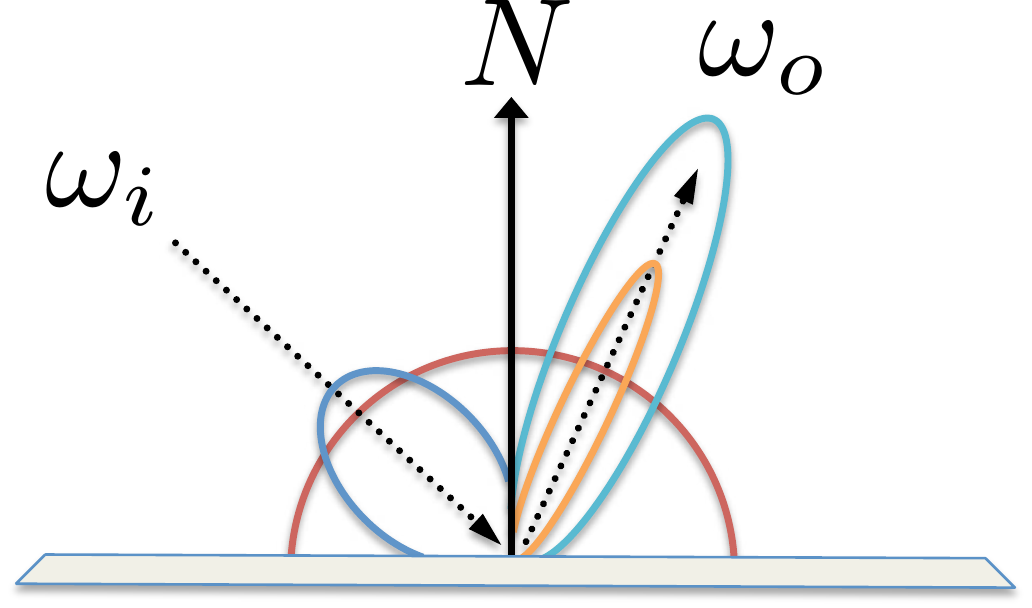} 
\includegraphics[width=0.49\columnwidth,page=2]{fig/parameterization/mat_rep.pdf} 
\vspace{-7.5mm} \\
{\color{white}'} \hspace{0.39\columnwidth} \fcolorbox{black}{white}{\Large a}  \hspace{0.42\columnwidth} \fcolorbox{black}{white}{\Large b}\vspace{1mm}\\
\includegraphics[width=0.49\columnwidth]{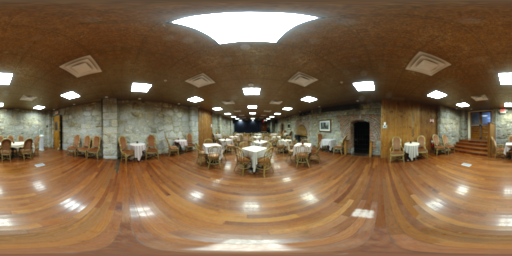}
\includegraphics[width=0.49\columnwidth]{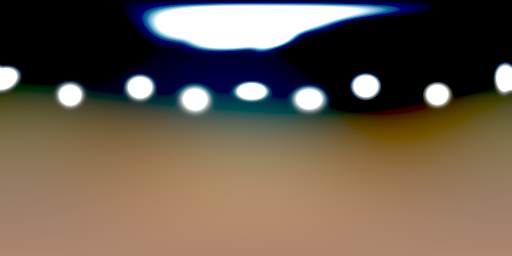} 
\vspace{-7.5mm} \\
{\color{white}'} \hspace{0.39\columnwidth} \fcolorbox{black}{white}{\Large c}  \hspace{0.42\columnwidth} \fcolorbox{black}{white}{\Large d}\vspace{0mm}\\
\centerline{\rule{.99\columnwidth}{0.2mm}}\\
\begin{minipage}{0.325\columnwidth}
\centerline{\scriptsize Measured BRDF (a)}\centerline{\scriptsize Environment map (c)}
\includegraphics[width=1\columnwidth]{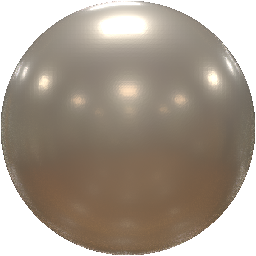}\\
\includegraphics[width=1\columnwidth]{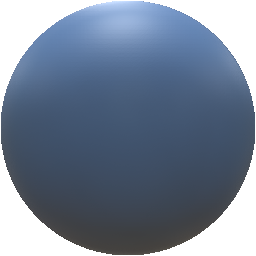}
\end{minipage}
\begin{minipage}{0.325\columnwidth}
\centerline{\scriptsize Our model (b)}\centerline{\scriptsize Environment map (c)}
\includegraphics[width=1\columnwidth]{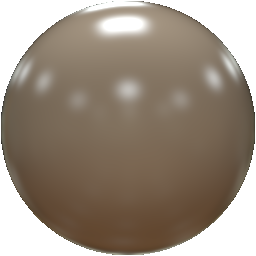}\\
\includegraphics[width=1\columnwidth]{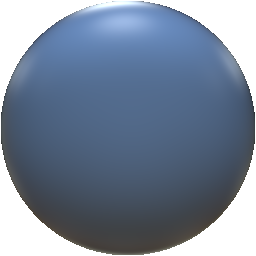}
\end{minipage}
\begin{minipage}{0.325\columnwidth}
\centerline{\scriptsize Our model (b)}\centerline{\scriptsize Our illumination (d)}
\includegraphics[width=1\columnwidth]{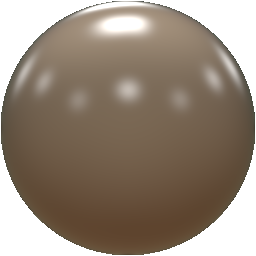}\\
\includegraphics[width=1\columnwidth]{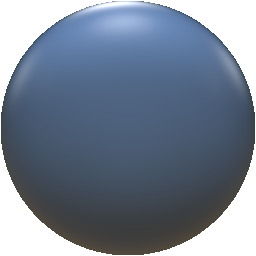}
\end{minipage}
\caption{A general BRDF can be made up of numerous reflection ``lobes'' (a), but in practice (e.g. surface modeling), a simple BRDF with one diffuse and one specular lobe tends to suffice (b). We use this representation as well as a low-order parameterization of illumination. Our illumination considers a real-world, omnidirectional lighting environment (c), and approximates it with a mixture of Gaussians and spherical harmonics; (d) shows our model fit to (c). We observe only slight perceptual differences when rendering with different combinations of the high- and low-order parameterizations (bottom rows). 
\vspace{0mm}
\vspace{0mm}
}
\label{fig:parameterization}
\end{figure}



\boldhead{Contributions} Our primary contribution is a technique for extracting spatially varying material reflectance (beyond diffuse parameters) directly from an object's appearance in a single photograph {\it without requiring any knowledge of the object's shape or scene illumination}. We use a low-order parameterization of material and develop a new model of illumination that can be described also with only a few parameters, allowing for efficient rendering. {\it Because our model has few parameters, we tend to get low variance and thus robustness in our material estimates (e.g. bias-variance tradeoff).} By design, our material model is the same that is used throughout the 3D art/design community, and describes a large class of real-world materials (Sec~\ref{sec:param}). We show how to efficiently estimate materials from plausible initializations of lighting and shape, and propose novel priors that are crucial in estimating material robustly (Sec~\ref{sec:est}). We extend this formulation to spatially varying materials in Section~\ref{sec:sv_est}. Our material estimates perform favorably to baseline methods and measure well with ground truth, and we demonstrate results for both synthetic and real images (Sec~\ref{sec:eval}). We show applications in relighting (Figs~\ref{fig:qualitative},~\ref{fig:natgeom}), material transfer (Fig~\ref{fig:teaser}), and material generation (Fig~\ref{fig:renderStealGrid}).

\begin{figure}[t]
\begin{minipage}{0.32\linewidth}  \centerline{ Input}
 \includegraphics[width=1\linewidth]{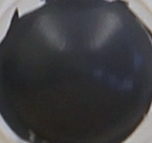} 
  \end{minipage}
\begin{minipage}{0.32\linewidth}  \centerline{Rendered}
 \includegraphics[width=1\linewidth]{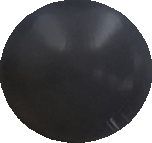}
 \end{minipage}
\begin{minipage}{0.32\linewidth} \centerline{ Mixing weights}
 \includegraphics[width=1\linewidth]{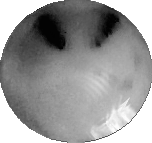} 
  \end{minipage} \\
  \centerline{\hrulefill}
  \\
\includegraphics[width=\linewidth]{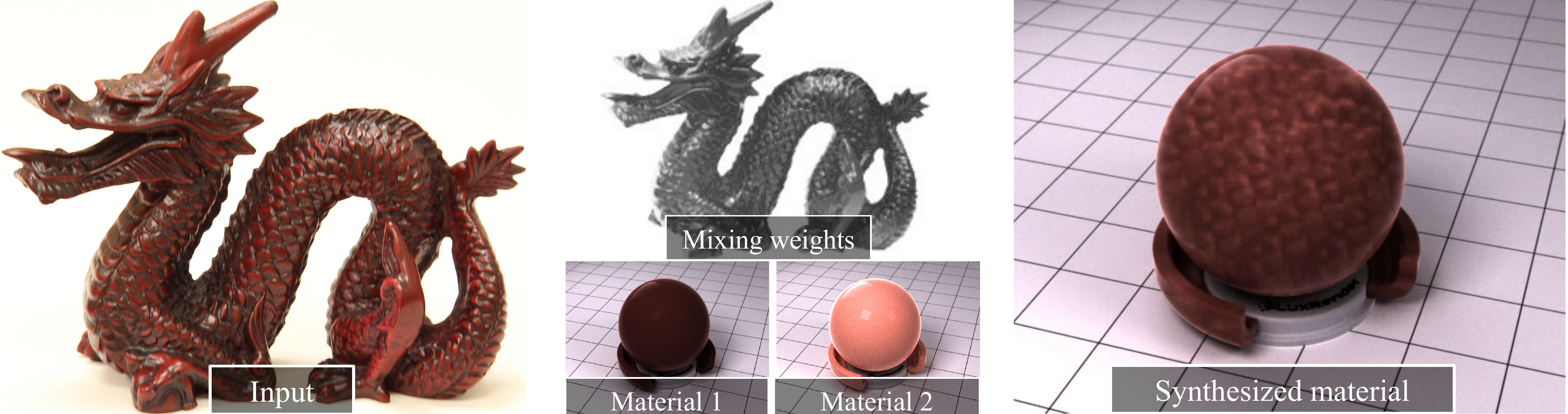}\\
\vspace{-6mm}
\caption{Failure examples. The top row demonstrates an incorrect mixture map estimate: specularities have been detected as a separate material. 
A material transfer result is shown on bottom, but our material model contains no mesostructure and appears flat.
}
\label{fig:failure}
\end{figure}

\boldhead{Limitations} Since we are using low-order material models that are isotropic and have monochromatic specular components, we cannot hope to estimate BRDFs of arbitrary shape (e.g. as measured by a gonioreflectometer), and there are some materials not encoded by our representation. Our recovered lighting and shape are not necessarily correct with respect to the true lighting/shape, although they are consistent with one another and sometimes give good estimates; as such, we only make claims about the accuracy of our material estimates. We use infinitely distant spherical lighting without considering interreflections, and we do not attempt to solve color constancy issues; lighting environments in our dataset integrate to the same value (per channel). Since we only have a single view of the object, certain material properties (e.g. specularities) may not be visible (depending especially on the coverage of the normals, i.e. flat surfaces provide much less information than curved surfaces). Due to our low-order model and perhaps mixture priors, shading effects can sometimes manifest in the spatial mixture map (Fig~\ref{fig:failure}). 

\section{Low-order reflectance and illumination} 
\label{sec:param}
Many previous methods have attempted to use high-order models of material (e.g. a linear combination of basis functions learned from measured BRDFs~\cite{Romeiro:eccv:10}) and illumination (parameterized with wavelets~\cite{Romeiro:eccv:10} or even on an image grid~\cite{Lombardi:eccv:2012,Oxholm:eccv:2012}, consisting of hundreds to thousands of parameters or more). We propose the use of more rigid models of shape and illumination, which still can describe the appearance of most real world objects, and provide necessary rigidness to estimate materials when neither shape or illumination are exactly known.

\boldhead{Representing material}
We represent materials using an isotropic diffuse and specular BRDF model consisting of only five parameters: diffuse albedo in the red, green, and blue channels ($R_d$), monochromatic specular albedo ($R_s$) and the isotropic ``roughness'' value ($r$), which is the concentration of light scattered in the specular direction, and can be considered (roughly) to be the size of the specular ``lobe'' (a smaller roughness value indicates a smaller specular lobe, where $r=0$ encodes a perfect specular reflector).

This type of material model is surprisingly general for its low number of parameters. Ngan et al. have previously shown that such parameterizations provide very good fits to real, measured BRDFs~\cite{ngan:egsr:05}.
Perceptually, this model can also encode a family of isotropic, dielectric surfaces (mattes, plastics, and many other materials in the continuum of perfectly diffuse to near-perfect specular)~\cite{PBRTv2}. There is also compelling evidence that such a material model suffices for photorealistic rendering, as this is the same material parameterization found most commonly in 3D modeling and rendering packages (such as Blender\footnote{\urlwofont{http://wiki.blender.org/index.php/Doc:2.6/Manual/Materials}}, which only considers diffuse and specular reflection for opaque objects), and used extensively throughout the 3D artist/designer community\footnote{\urlwofont{http://www.luxrender.net/forum}}.

We write our BRDF following the isotropic substrate model as described in {\it Physically Based Rendering}~\cite{PBRTv2}, which uses a microfacet reflectance model and assumes the Schlick approximation to the Fresnel effect~\cite{Schlick94aninexpensive}. 

Figure~\ref{fig:parameterization} shows a comparison of what a measured BRDF (a) might look like in comparison to our material parameterization (b). We compare measured BRDFs to our material model (fit using the procedure described in Sec~\ref{sec:eval}) rendered in natural illumination in the bottom row (left two columns). 

\boldhead{Representing illumination}
Consider a single point within a scene and the omnidirectional light incident to that point. This incident illumination can be conceptually decomposed into luminaires (light-emitters) and non-emitting objects.  We consider these two separately, since the two tend to produce visually distinct patterns in object appearance (depending of course on the material). Luminaires will generally cause large, high-frequency changes in appearance (e.g. specular highlights), and non-emitters usually produce small, low-frequency changes. 

Using this intuition, we parameterize each luminaire as a two dimensional Gaussian in the 2-sphere domain (sometimes known as the Kent distribution), and approximate any other incident light (non-emitted) as low-order functions on the sphere using $2^{nd}$ order spherical harmonics\footnote{We assume all lighting comes from an infinite-radius sphere surrounding the object, as done in previous methods~\cite{Barron:eccv:12,Lombardi:eccv:2012,Oxholm:eccv:2012,Romeiro:eccv:10})}. 

Such a parameterization has very few parameters relative to a full illumination environment (or environment map): six per light source (two for direction $L^{(d)}$, one for each intensity $L^{(I)}$, concentration $\kappa$, ellipticalness $\beta$, and rotation about the direction $\gamma$) and 27 spherical harmonic coefficients (nine per color channel), but more importantly, this parameterization still enables realistic rendering at much higher efficiency. Rendering efficiency is crucial to our procedure, as each function evaluation in our optimization method (Sec~\ref{sec:est}) requires rendering.

Our lighting environments maintain only high frequencies in regions of emitting sources, and is encoded by low-frequency spherical harmonics everywhere else. However, rendering with full versus approximate (our) lighting produces similar results (bottom middle vs bottom right). For additional discussion, see Sec~\ref{sec:est}.

\begin{figure*}
\centerline{\large True \hspace{27mm} Starting point \hspace{17mm} Optimized estimates \hspace{25mm} Final result}
\hspace{2mm} Input image \hspace{7mm} Normals \hspace{7mm} Rendered \hspace{9mm} Normals  \hspace{7mm} Rendered \hspace{8mm} Normals \hspace{7mm} True ${\bf M}$ (input) \hspace{5mm} Recovered ${\bf M}$  \hspace*{2mm} \\
\setlength\fboxrule{1pt}
\setlength\fboxsep{1pt}
\begin{overpic}[width=\linewidth]{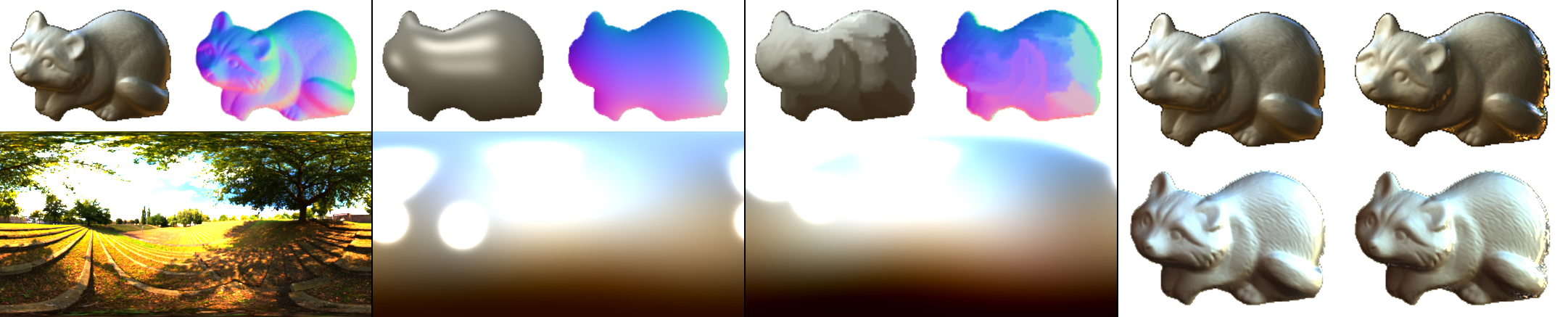}
\put(391,53){\fcolorbox{black}{white}{\footnotesize Original illumination}}
\put(395,0){\fcolorbox{black}{white}{\footnotesize Novel illumination}}
\end{overpic}
\vspace{-4mm}
\caption{Results from our optimization procedure. On the left is the input image (top left), true surface normals (top right), and true illumination (below), followed to the right by estimates that we begin our optimization with. The estimated rendering (using Eq~\ref{eq:render}), estimated surface normals, and estimated illumination are displayed in the third column. The rightmost column shows the true material (left) and our estimated material (right) rendered onto the true shape in the original lighting environment (top), and rendered in novel lighting (bottom). Our initialization is described in Sec~\ref{sec:init}, and uses no prior information about the input.
\vspace{-4mm}
}
\label{fig:optimization}
\end{figure*}

\section{Estimating specular reflectance}
\label{sec:est}
Our idea is to jointly recover material, shape, and illumination, such that a rendering of these estimates produces an image that is similar to the input image. Following Barron and Malik~\cite{Barron:eccv:12}, we also enforce a strong set of priors to bias our material estimate towards plausible results. 



Our goal is to recover a five dimensional set of material parameters $\ve{M} = (R_d^{(r)},R_d^{(g)},R_d^{(b)}, R_s,r)$, while jointly optimizing over illumination $\ve{L}$ and surface normals $\ve{N}$. Following notation in section~\ref{sec:param}, we denote $R_d$ as RGB diffuse reflectance, $R_s$ as monochromatic specular reflectance, and $r$ as the roughness coefficient (smaller $r \Rightarrow$ narrower lobe $\Rightarrow$ shinier material). Illumination $\ve{L} = \{L,s\}$ is parameterized as a mixture of $m$ Gaussians in the 2-sphere domain $L = \{L_1,\ldots L_k\}$ with direction $L^{(d)}_i$, intensity $L^{(I)}_i$, concentration $L^{(\kappa)}_i$, ellipticalness $L^{(\beta)}_i$, and rotation $L^{(\gamma)}_i$ ($i \in \{1,\ldots,k\}$). Statistics of real illumination environments are nonstationary and can contain concentrated bright points~\cite{dror2004sta}; the Gaussian mixture aim to represent these peaks. Indirect light $s$ is represented as a 9 $\times$ 3 matrix of $2^{nd}$ order spherical harmonic coefficients (9 per color channel). $\ve{N}$ is simply a vector of per-pixel surface normals parameterized by azimuth and elevation directions.

We phrase our problem as a continuous optimization by solving the parameters which minimize the following:
\begin{align}
\label{eq:obj}
\hspace{5mm} \displaystyle \underset{{\ve{M},\ve{N},\ve{L}}}{\operatorname{argmin}} \ \ &E_{\text{rend}}(\ve{M},\ve{N},\ve{L} ) + E_{\text{mat}}(\ve{M}) \ + \nonumber \\ 
  &E_{\text{illum}}(\ve{L} ) + E_{\text{shape}}(\ve{N})  \nonumber  \\ \nonumber \\ 
\text{subject}&\text{ to } 0 \leq \ve{M}^{(i)} \leq 1,\ i\in\{1,\ldots,5\},
\end{align} \vspace{-.5mm}
where $E_{\text{rend}}$ is the error between a rendering of our estimates and the input image, and $E_{\text{mat}},E_{\text{illum}}$, and $E_{\text{shape}}$ are priors that we place on material, illumination, and shape respectively. In the remainder of this section, we discuss the rendering term and the prior terms. Figure~\ref{fig:optimization} shows the result of our optimization technique at various stages for a given input.

\boldhead{Rendering error} Our optimization is guided primarily by a term that penalizes pixel error between the input image and a rendering of our estimates. The term itself is quite simple, but efficiently optimizing an objective function (which includes the rendering equation) can be challenging. Writing $I$ as the input, we define the term as the average squared error for each pixel: 
\begin{equation}
E_{\text{rend}}(\ve{M},\ve{N},\ve{L} ) = \sum_{i\in \text{pixels}} \sigma_i^{rend} || I_i - f(\ve{M},\ve{N}_i,\ve{L} ) ||^2,
\end{equation}
where $f(\ve{M},\ve{N},\ve{L} )$ is our rendering function, and $\sigma_i^{rend} =  I_i^2$ re-weights the error to place more importance on brighter points (primarily specularities).

Notice that we do not strictly enforce equality (as in~\cite{Barron:eccv:12,Barron:cvpr:12}), as this soft constraint allows more flexibility during the optimization, and because our parameterizations are too stiff for equality to hold. This has the added benefit of reducing variance in our estimates.

As in any iterative optimization scheme, each iteration requires a function evaluation (and most likely a gradient or even hessian depending on the method). If chosen na\"{i}vely, $f$ can take hours or longer to evaluate and differentiate, and here we describe how to construct $f$ so that this optimization becomes computationally tractable.

The key to efficiency is in our low-order parameterization of illumination. By considering emitting and non-emitting sources separately, we treat our rendering as two sub-renders: one is a ``full'' render using the emitting  luminaires (which are purely directional since we assume the light is at infinity), and the other is a diffuse-only render using all other incident light (reflected by non-emitters).

Denoting $\Omega_e$ and $\Omega_n$ as set of ``emitting'' and ``non-emitting'' light directions respectively, $l(\omega)$ as the light traveling along direction $\omega$, and $f_M$ as the BRDF defined by material $M$, we write our rendering function as
\begin{align}
f^{(e)}_i &=  \int_{\Omega_e} f_M(\omega,v) l(\omega) \max(\omega \cdot \ve{N}_i,0) d\omega \nonumber \\
f^{(n)}_i &=   \int_{\Omega_n} l(\omega) \max(\omega \cdot \ve{N}_i,0) d\omega \nonumber \\
f(\ve{M},\ve{N},\ve{L} )_i &= f^{(e)}_i + R_d f^{(n)}_i,
\label{eq:render}
\end{align}
for the $i^{th}$ image pixel and a particular view direction $v$. Notice that $f^{(n)}_i$ is simply irradiance over the non-emitting regions of the sphere, and is modulated by diffuse reflectance ($R_d$) since Lambertian BRDFs are constant.

We can compute both of these efficiently, because $\Omega_e$ is typically small (most lighting directions are occupied by negligible Gaussian components), and it is well known that diffuse objects can be efficiently rendered through spherical harmonic projection~\cite{Ravi:sg:2001}.
In terms of previous notation, directional sources $L$ are used in the full render ($f^{(e)}_i$), and $s$ is used for the diffuse-only render ($f^{(n)}_i$). 

The intuition behind such a model is that indirect light contributes relatively low-frequency effects to an object's appearance, and approximating these effects leads to only slight perceivable differences~\cite{Ravi:sg:01b}.  A variation of this intuition is used for efficiently choosing samples in Monte Carlo ray tracing (e.g. importance sampling~\cite{PBRTv2}), which causes problems in continuous optimization techniques since rendering is then non-deterministic. 


\boldhead{Material prior} The rigidity of our material model (5 parameters to describe the entire surface), is a strong implicit prior in itself, but we also must deal with the ambiguity that can exist between diffuse and specular terms. For example, if a specular lobe ($r$) is large enough, then the specular albedo and diffuse albedo can be confused (e.g. dark specular albedo/bright diffuse albedo may look the same as bright specular albedo/dark diffuse albedo). Thus, we add a simple term to discourage large specular lobes, persuading the diffuse component to pick up any ambiguity between it and the specular terms:
\begin{equation}
E_{\text{mat}}(\ve{M}) = \lambda_m r^2.
\label{eq:matprior}
\end{equation}
The only material parameter that is constrained is specular lobe size, and $\lambda_m = 1$ in our work. 

\boldhead{Illumination prior}
We develop our illumination prior by collecting statistics from spherical HDR imagery found across the web (more details in Sec~\ref{sec:eval}). Each image gives us a sample of real-world illumination, and to see how each sample relates to our illumination parameters, we fit our lighting model (SO(2) Gaussians + 2${}^{nd}$ order spherical harmonics) to each spherical image. Fitting is done using constrained, non-linear least squares, and the number of Gaussians (corresponding roughly to luminaires) is determined by the number of peaks in the HDR image (smoothed to suppress noise). Priors are developed by clustering the Gaussian parameters, and through principal component analysis on the spherical harmonic coefficients.

Denote $\bar\kappa_j$, $\bar\beta_j$ as the means of the $j^{th}$ clusters (clustered independently using $k$-means) for the concentration, and ellipticalness of Gaussian parameters from our fitting process. Intuitively, these cluster centers give a reasonable sense of the shape of luminaires found in typical lighting environments, and we enforce our estimated sources to have shape parameters similar to these:
\begin{equation}
E_{\text{illum}}^{\text{means}}(L_i) = \mathcal{S}( \{ |L^{(\kappa)}_i - \bar\kappa_j| \}_{i=1}^k ) + \mathcal{S}( \{ |L^{(\beta)}_i - \bar\beta_j| \}_{i=1}^k ),
\end{equation}
where $\mathcal{S}$ is the softmin function (differentiable min approximation) and $|\cdot|$ is a differentiable approximation to the absolute value (e.g. $\sqrt{x^2+\epsilon}$). 

We also find the principal components (per channel) of the spherical harmonic coefficients fit to our data. During estimation, we reparameterize the estimated SH coefficients using weight vectors $w_{\{r,g,b\}}$, principal component matrices $S_{\{r,g,b\}}$, and means of all fit SH components $\mu_{\{r,g,b\}}$: $s(w) = [\mu_r + S_r w_r, \mu_g + S_g w_g, \mu_b + S_b w_b]$. We impose a Laplacian prior on the weight vector:
\begin{equation}
E_{\text{illum}}^{\text{pca}}(w) = \sum_{i\in\text{weights}} | w |.
\end{equation}
This coerces the recovered SH components to lie near the dataset mean, and slide along prominent directions in the data. We found that seven principal components (per channel) roughly explained over 95\% of our data (eigenvalue sums contain $>$95\% of the mass), and we discard the two components corresponding to the smallest eigenvalues (then $w_{\{r,g,b\}} \in \mathbb{R}^7$ and $S_{\{r,g,b\}} \in \mathbb{R}^{9 \times 7}$).

We also impose a gray world assumption, namely that each color channel should integrate (over the sphere of directions) to roughly the same value. Because we only have a single view of an object, some portions of the lighting sphere have significantly more influence than others; e.g. the hemisphere behind the object is mostly unseen and has smaller influence than the hemisphere in front. We weight the integration appropriately so that the dominant hemisphere has more influence (using $W_\theta = \cos(\frac{\theta-\pi}{2})$, where $\theta$ is the angle between the view direction and the direction of integration). This integration translates to a simple inner product (due to the nice properties of spherical harmonics), making the prior easy to compute:
\begin{eqnarray}
E_{\text{illum}}^{\text{gray}}(s) = & || G^T s_r - G^T s_g || + || G^T s_g - G^T s_b || \nonumber \\
 &+ || G^T s_r - G^T s_b ||,
\end{eqnarray}
where $G$ is the pre-computed integral of $2^{nd}$ order spherical harmonic basis functions (weighted by $W_\theta$), and $s_{\{r,g,b\}}$ are the current estimates of spherical harmonic coefficients.

Our prior is then a weighted sum of these three terms:
\begin{equation}
E_{\text{illum}}({\bf L}) = \lambda_m \sum_{i=1}^m E_{\text{illum}}^{\text{means}}(L_i) + \lambda_p E_{\text{illum}}^{\text{pca}}(w) + \lambda_g  E_{\text{illum}}^{\text{gray}}(s),
\end{equation}
keeping in mind ${\bf L} = \{L,s(w)\}$, and $w$ as PCA weights described above. We set $\lambda_m = \lambda_p = \lambda_g = 0.1$.

\boldhead{Shape prior}
We also optimize over a grid of surface normals, and impose typical shape-from-contour constraints: smoothness, integrable shape, and boundary normals are assumed perpendicular to the view direction. Let $N_i = (N_i^x, N_i^y, N_i^z)$, and $\hat N$ be the set of normals perpendicular to the occluding contour and view direction. We write the prior as:
\begin{eqnarray}
E_{\text{shape}}({\bf N}) =& \displaystyle\sum_{i \in\text{pixels}} \lambda_s \eta_i^{s} || \nabla N_i || + \lambda_I ||\nabla_y \frac{N^x_i}{N^z_i} -  \nabla_x \frac{N^y_i}{N^z_i} || \nonumber \\ 
&+ \lambda_{c} \sum_{c \in\text{boundary pixels}} ||N_c - \hat N_c||, 
\end{eqnarray}
where the first term encodes smoothness where the input is also smooth (modulated by image dependent weights $\eta$), the second term enforces integrability, and the third ensures that boundary normals are perpendicular to the viewing direction. We set the weights as $\lambda_s=1$, $\lambda_I=\lambda_c=0.1$.

\subsection{Initialization}
\label{sec:init}
Initial estimates of shape come from a na\"{i}ve shape-from-contour algorithm (surface assumed tangent to view direction at the silhouette, and smooth elsewhere), and light is initialized with the mean of our dataset (if applicable; leaving out the illumination that generated the input image). We estimate an initial $R_d$ by rendering irradiance with initial estimates of shape and lighting, dividing by the input image to get per-pixel albedo estimates, and averaging the RGB channels; $R_s, r$ are set as small constants (0.01 for our results). Full details of our initialization procedure can be found in supplemental material.

\subsection{Undoing estimation bias} 
\label{sec:bias}
Our low-parametric models tend to introduce bias into our estimates, but at the same time reduce estimation variance; e.g. bias-variance tradeoff). However, we have found that our priors produce consistent estimation bias: we typically see a smaller specular lobe and specular albedo due most likely to our material prior (Eq~\ref{eq:matprior}). We may also observe omitted-variable bias for images with materials not encoded by our model, but we do not address here.

Past methods point out that there are clear visual distinctions between different types and levels of gloss~\cite{fleming2003rea,Sharan:josaa:2008,Wills:tog:2009}, and we use the input image coupled with our estimates to develop an ``un-biasing'' function. We develop a simple regression method (simple methods should suffice since the bias appears to be consistent) which works well for removing bias and produces improved results.  Our goal is to find a linear prediction function that takes a vector of features to unbiased estimates of $R_d, R_s$ and $r$. Our features consist of our estimates of specular albedo and specular lobe size, as well as histogram features computed on the resulting rendered image, normal map, input image, and the error image (rendered minus input); features are computed for both raw and gradient images. Given a set of results from our optimization technique with ground truth material parameter (obtained, e.g., from our dataset in Sec~\ref{sec:eval}), we compute a bias prediction function by solving an L${}_1$ regression problem (with L${}_2$ regularization). For more details, see the supplemental material.

%
%

\section{Recovering spatially varying reflectance}
\label{sec:sv_est}

We propose an extension of Eq~\ref{eq:obj} for estimating spatial mixtures of materials. First, we define our appearance model simply as a spatially varying linear combination of renderings. Radiance at pixel $i$ is defined as:
\begin{equation}
\sum_{j\in \text{materials}} m_{i,j} f(\ve{M}_j,\ve{N}_i,\ve{L} ),
\label{eq:apperancemodel}
\end{equation}
where $m_{i,j}$ is the $j^{th}$ mixture weight at pixel $i$, and $\ve{M}_j$ is the $j^{th}$ material. The rendering error term for estimating spatial materials then becomes:
\begin{eqnarray}
 E_{\text{rend}}^{\text{mix}}(\ve{M}_1,\ldots,\ve{M}_k,\ve{N},\ve{L}, m) =  & \nonumber \\
 \hspace{-2mm} \sum_{i\in \text{pixels}} \sigma_i^{rend} \Big|\Big| I_i -  \hspace{-4mm} \displaystyle\sum_{j\in \text{materials}} m_{i,j} & f(\ve{M}_j,\ve{N}_i,\ve{L} ) \Big|\Big|^2.
\end{eqnarray}

We define three properties that the spatial maps ($m$) must adhere to: unity, firmness\footnote{We define the firmness prior as the decisiveness of mixture weights to snap to 0 or 1, and in this sense it has no relation to tactile properties.}, and smoothness. First, the unity prior ensures that the mixture weights must be nonnegative and sum to one at every pixel:
\begin{equation}
 \forall i,j \ \ m_{i,j} > 0, \hspace{5mm} \forall i \ \ \sum_{j} m_{i,j} = 1.
\end{equation} 
As noted by Goldman et al.~\cite{Goldman:tpami:2010}, this prevents overfitting and removes certain ambiguities during estimation.

We place another prior on the ``firmness'' of our mixture maps. For certain objects, many patches on the surface are dominated by a single material (e.g. checkerboard); for others, the surface is roughly uniform over space (e.g. soap can be made of a diffuse layer and a glossy film which are both present over the whole surface); there are even materials ranging in between (e.g. marble). We would like a structured way of controlling which type of spatial mixture we produce, and we do so by imposing an exponential prior on each mixture element:
\begin{equation}
E_{\text{firm}}^{\text{mix}}(m) = \sum_{i,j} m_{i,j}^\alpha,
\label{eq:firm}
\end{equation}
where $\alpha > 0$ controls how firm a mixture will be. For example, with the unity constraint, $\alpha>1$ encourages uniform mixture weights (not firm, e.g. soap), and  $\alpha<1$ encourages mixture weights to be near zero or one (firm, e.g. checkerboard). For results in this paper, we use $\alpha=0.5$. Notice that for $\alpha<1$ this function is no longer convex, although in practice our optimization still seems to fare well. 
Our prior is more general (and controllable) than the method of Goldman et al.~\cite{Goldman:tpami:2010}, which assumes that each pixel is the linear combination of {\it at most} two materials.

Finally, we encourage spatial smoothness of the mixtures, as nearly all mixed-materials contain spatial structure:
\begin{equation}
E_{\text{smooth}}^{\text{mix}}(m) = \sum_{i,j} ||\nabla_x m_{i,j}|| + ||\nabla_y m_{i,j}||,
\label{eq:smoothness}
\end{equation}
where $\nabla_x$ and $\nabla_y$ are spatial gradient operators in the image domain. 

By inserting our new rendering term and mixture priors into the objective function for single materials (Eq~\ref{eq:obj}), we define a new optimization problem for estimating spatially varying materials:
\begin{align}
 \hspace{0mm} \displaystyle \underset{{\ve{M}_1,\ldots,\ve{M}_k,\ve{N},\ve{L}}}{\operatorname{argmin}} \ \ & E_{\text{rend}}^{\text{mix}}(\ve{M}_1,\ldots,\ve{M}_k, \ve{N},\ve{L},m) +  E_{\text{mat}}(\ve{M})  +\nonumber \\
& \hspace{-10mm} E_{\text{illum}}(\ve{L} ) + E_{\text{shape}}(\ve{N}) + E_{\text{firm}}^{\text{mix}}(m)  + E_{\text{smooth}}^{\text{mix}}(m),  \nonumber \\
& \hspace{-10mm}  \text{subject to:    \ \ \ } 0 \leq \ve{M}_j^{(i)} \leq 1,\ i\in\{1,\ldots,5\}, \ \forall j, \nonumber \\
& \hspace{8.5mm}  \forall i,j \ \ m_{i,j} > 0, \nonumber \\ 
& \hspace{8.5mm} \forall i \ \ \sum_{j} m_{i,j} = 1.
\label{eq:obj_mix}
\end{align}
Solving this objective function can be difficult, but we have had success using constrained quasi-Newton methods (L-BFGS Hessian).

Our optimization results in a decomposition of the input image into $k$ materials $\ve{M}$, a set of per-pixel weights for each material $m$, per-pixel surface normals $\ve{N}$, and illumination parameters $\ve{L}$. In this work, we focus on the correctness of our mixture materials and their applications.

\begin{figure*}[t]
\includegraphics[width=.32\linewidth]{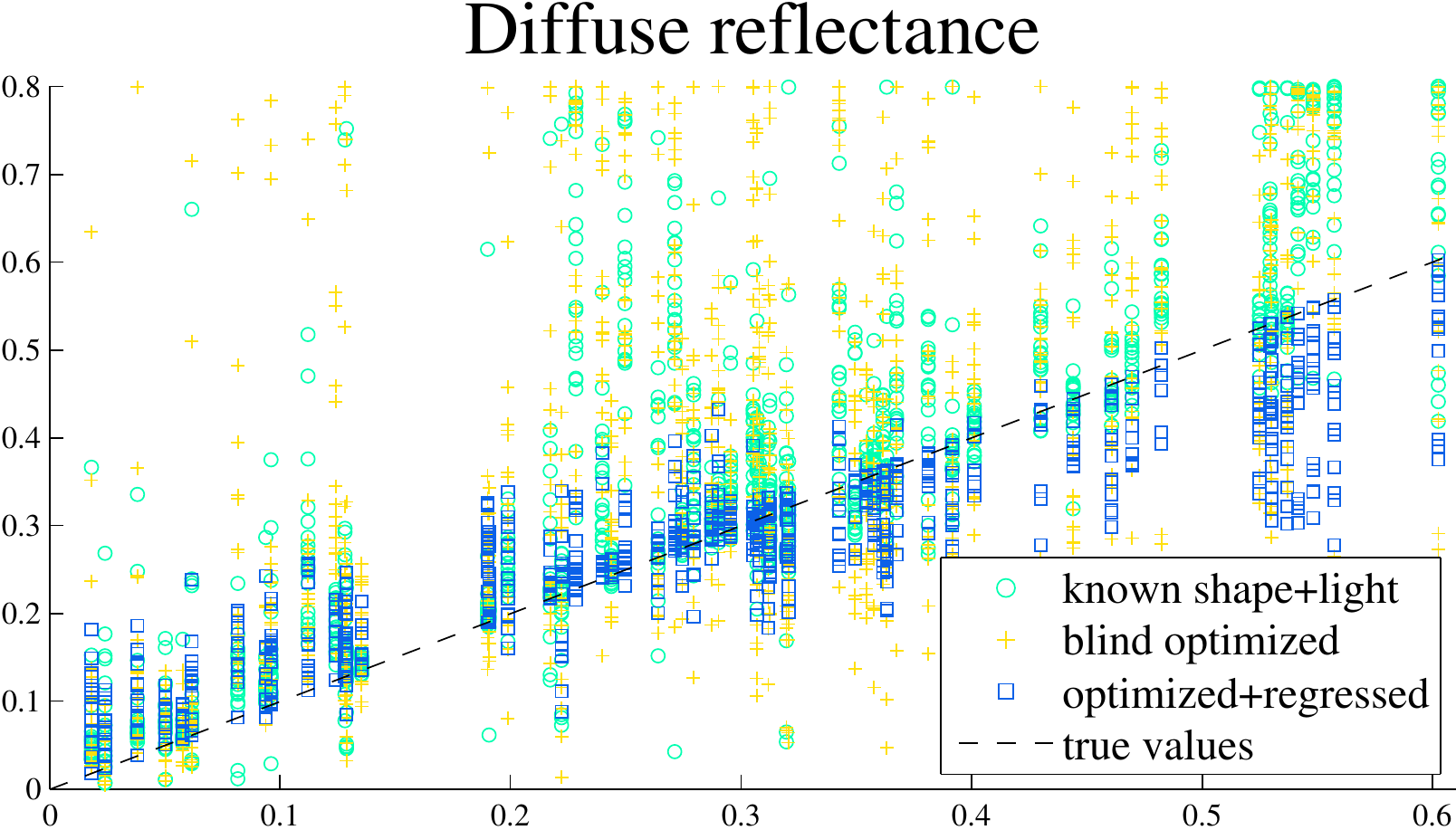}
\includegraphics[width=.32\linewidth]{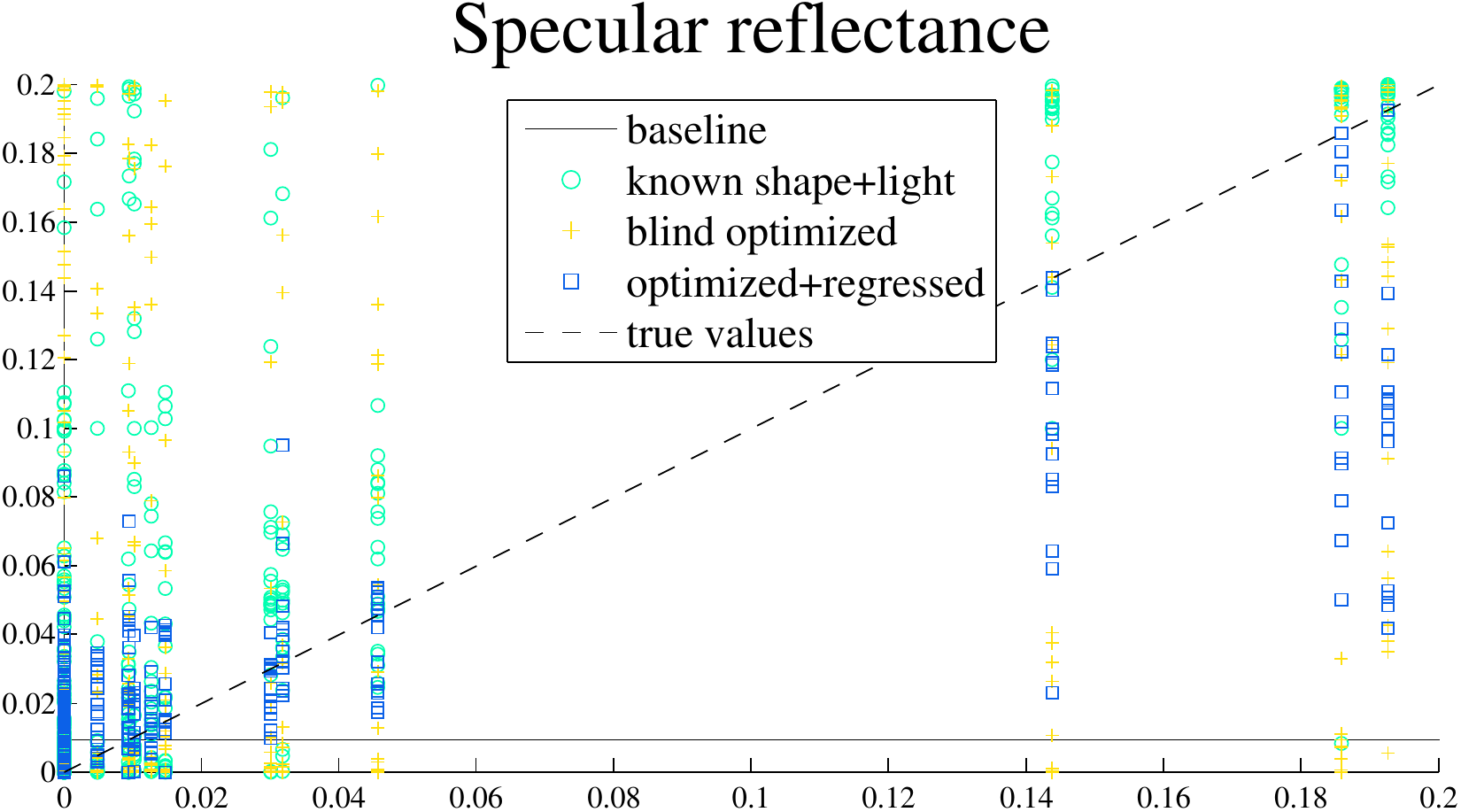}
\includegraphics[width=.32\linewidth]{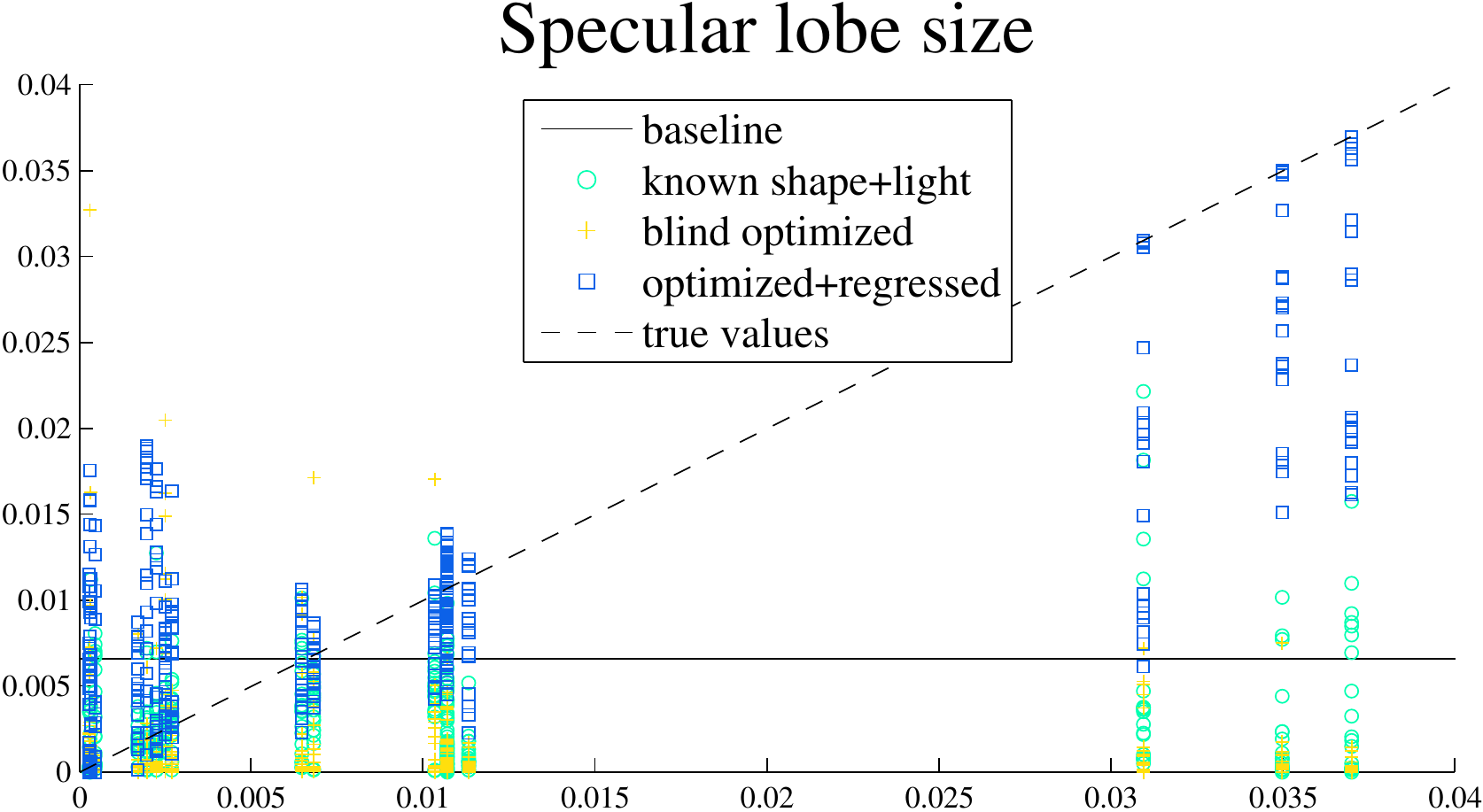}
\caption{Errors in material estimates for each image in our dataset. Each plot shows the true material value on the horizontal axis plotted against our estimate of diffuse reflectance ($R_d$), specular reflectance $R_s$, and specular lobe size $r$ (left to right). We show the results for our baseline, the material produced given accurate initial shape and lighting, our blind optimization technique (blind optimized), and the material regressed by un-biasing our optimization results (blind regressed); details in Sec~\ref{sec:eval}.
\vspace{-1mm}
}
\label{fig:mat_err}
\end{figure*}

\begin{figure*}[t]
\centerline{
\includegraphics[width=0.45\linewidth]{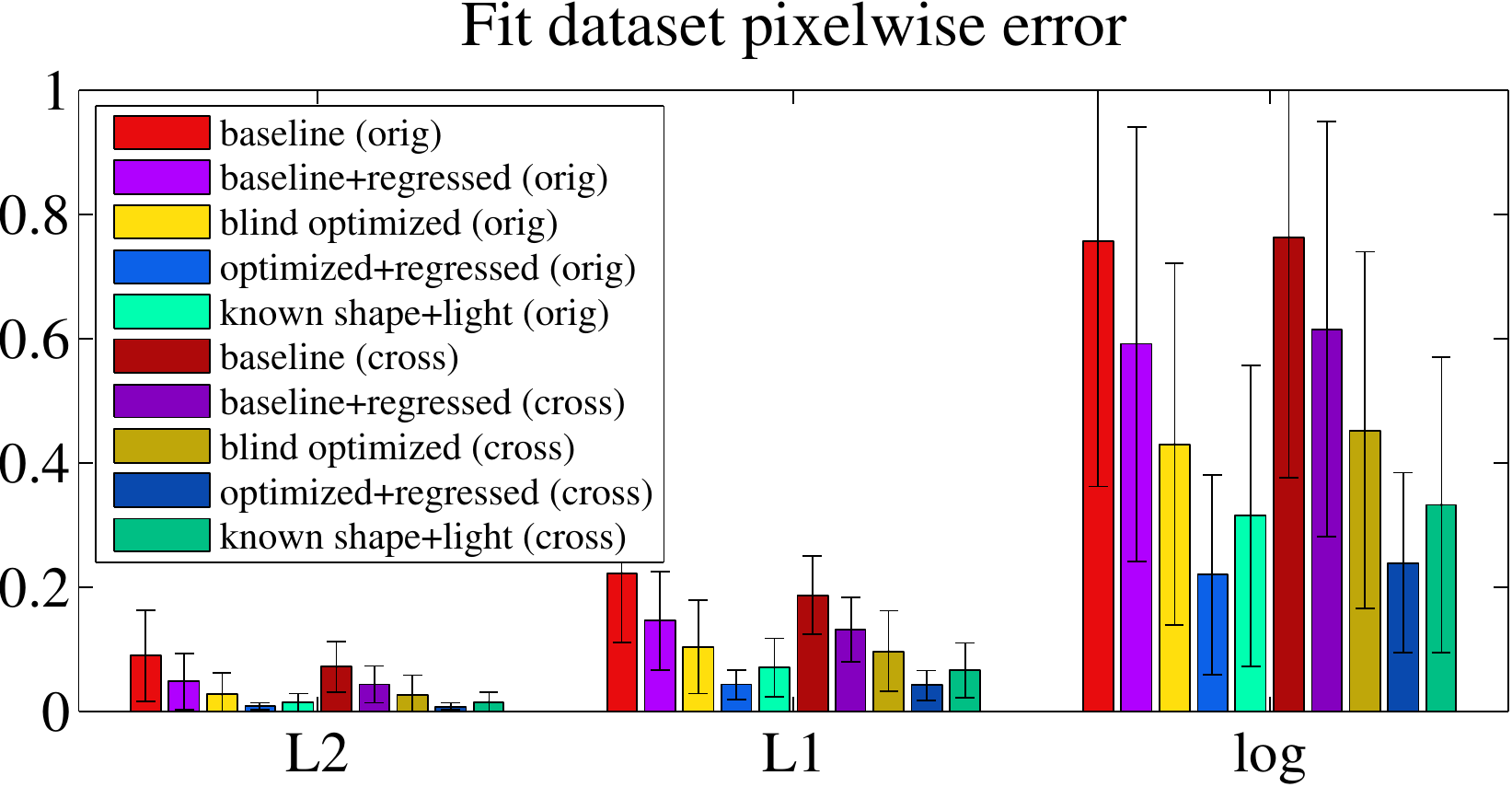}
\hspace{5mm}
\includegraphics[width=0.45\linewidth]{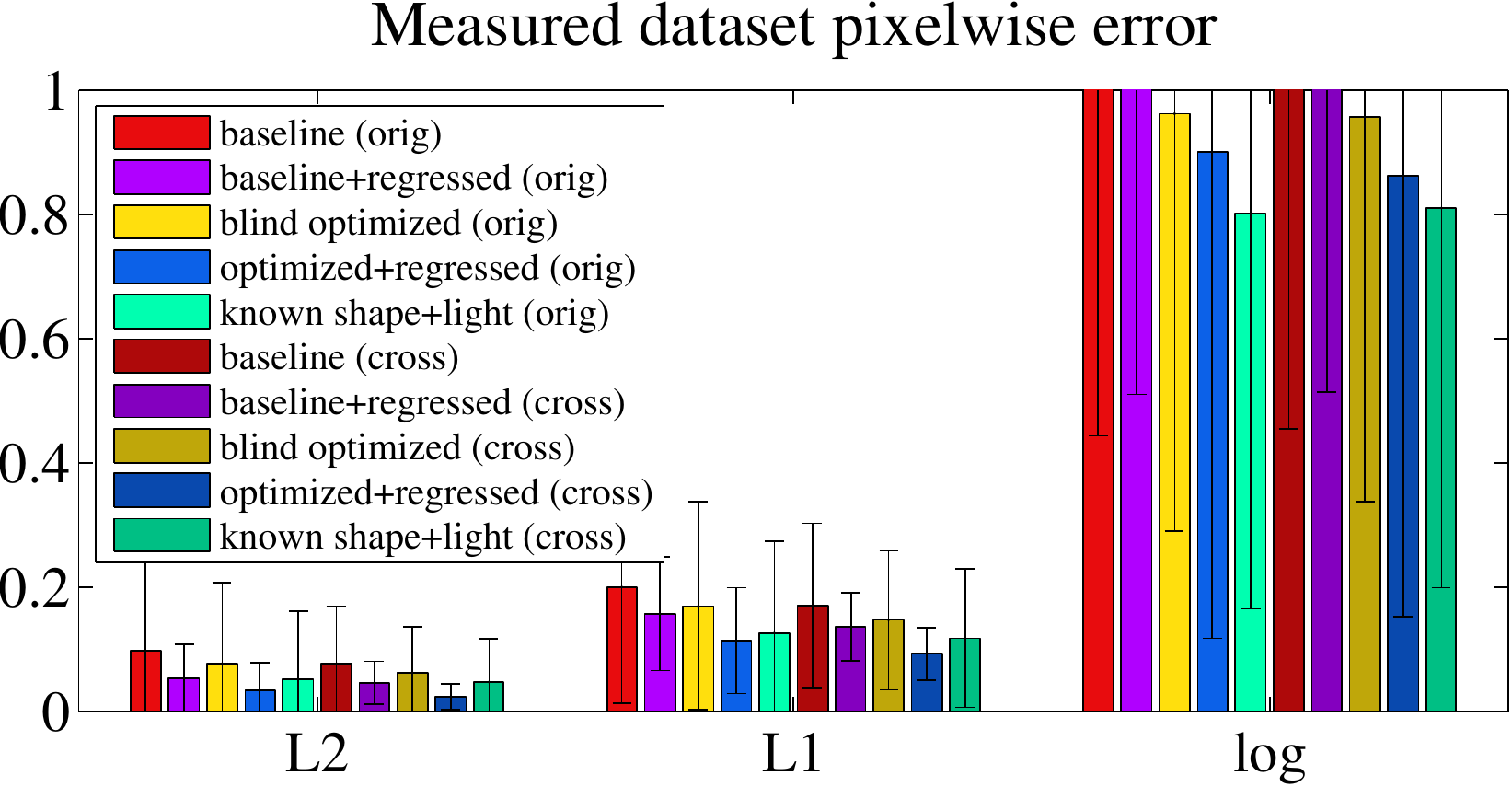}
\vspace{-1mm}
}
\caption{We compare the average per-pixel error of the input image and a re-rendered image with estimated material ({\it but with the true shape and true lighting that produced the input image}) for various techniques and for both versions of our dataset; see Sec~\ref{sec:eval} for details. We compute errors in the original illumination (orig), and averaged over six novel illumination environments (cross), for three different metrics: L2 and L1 norm, and the absolute log difference, and show the mean over the dataset (error bars indicate one standard deviation). Our full method (optimized+regressed) achieves low error relative to others. We also observe similar (yet slightly worse) error on our measured dataset, indicating that, for a variety of cases, our a) our method can handle real-world materials, and b) that our material model is capable of visually reproducing complex reflectance functions.
\vspace{-2mm}
}
\label{fig:rendererror}
\end{figure*}

\section{Experiments}
\label{sec:eval}

We evaluate the results of our method for objects with homogeneous (spatially uniform) materials in Section~\ref{sec:eval:uniform}, as well as our inhomogeneous (spatially varying) material results in Section~\ref{sec:eval:sv}. We report results for both a dataset we collected containing ground truth material information, as well as for the Drexel Natural Illumination dataset.

\subsection{Homogeneous materials}
\label{sec:eval:uniform}

For evaluation and training our bias predictors (Sec~\ref{sec:bias}), we have collected a dataset consisting of  400 images rendered with real world shapes, materials, and illumination environments (all chosen from well-established benchmark datasets). We use the 20 ground truth shapes available in the MIT Intrinsic Image dataset~\cite{Grosse:iccv:09}, and render each of these objects with 20 of the materials approximated from the MERL BRDF dataset~\cite{Matusik:sg:03}, for a total of 400 images. We use 100 different illumination environments (50 indoor, 50 outdoor) found across the web, primarily from the well known ICT light probe gallery\footnote{\urlwofont{http://gl.ict.usc.edu/Data/HighResProbes}} and the sIBL archive\footnote{\urlwofont{http://www.hdrlabs.com/sibl/archive.html}}. We ensure that each object is rendered in 10 unique indoor and 10 unique outdoor lighting environments, permuted such that each illumination environment is used exactly four times throughout the dataset. Each lighting environment is white balanced and has the same mean (per channel).

Our dataset has two ``versions.'' The first version of our dataset ({\bf fit dataset}) is rendered using our low-order reflectance model (we approximate MERL BRDFs by fitting our own 5-parameter material model to the measured data, and render using our fits). The resulting images are highly realistic, and allow us to both compare our material estimates with ground truth, and regress bias prediction functions (as in Sec~\ref{sec:bias}).

The second version ({\bf measured dataset}) is rendered using only {\it measured} BRDFs (from the MERL dataset); these images are truly realistic as the shape, material, and lighting are all sampled directly from real-world data. Furthermore, these images are synthesized using a physical renderer and thus include shadows and bounced light. This dataset gauges how well our method can generalize to real images and reflectances not encoded by our model.

\begin{figure*}
{\large \hspace{39mm} \underline{Fit dataset} \hspace{60mm} \underline{Measured dataset}}

{ \hspace{26mm} Best results \hspace{12mm} Median results \hspace{30mm} Best results \hspace{10mm} Median results \hspace{10mm} }\\
\begin{minipage}{.09\linewidth}
\hfill
  \begin{sideways}{\large \underline{Novel illumination} \hspace{4mm} \underline{Original illumination}}\end{sideways}
  \begin{sideways}{ \hspace{-5mm} true \hspace{2mm} opt+reg \hspace{4mm} opt \hspace{8mm}  true \small{(input)} \hspace{1mm} opt+reg \hspace{4mm} opt}\end{sideways}
  \hspace{2mm}
\end{minipage}
\begin{minipage}{.45\linewidth}
\includegraphics[height=\linewidth]{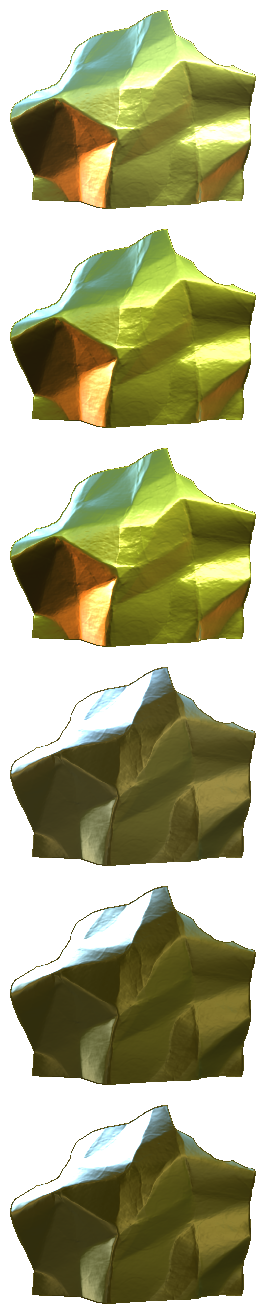}
\includegraphics[height=\linewidth]{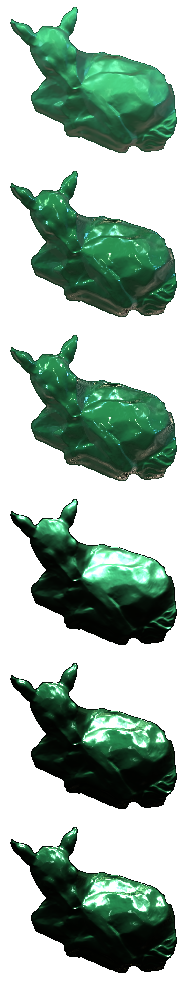}
\includegraphics[height=\linewidth]{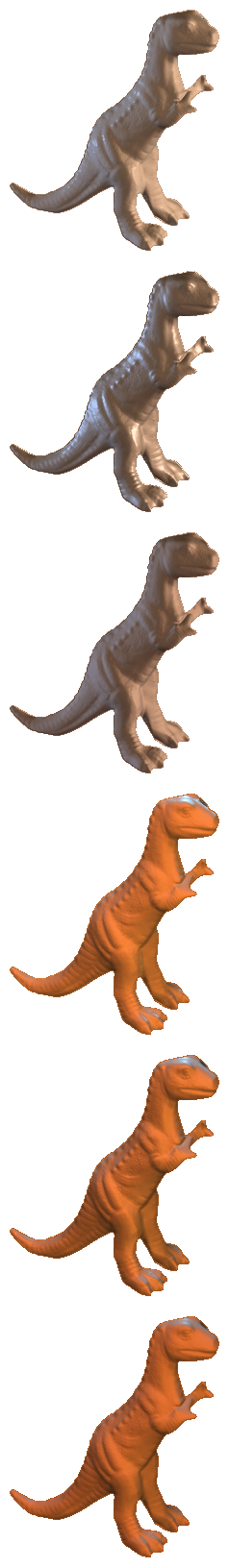}
\includegraphics[height=\linewidth]{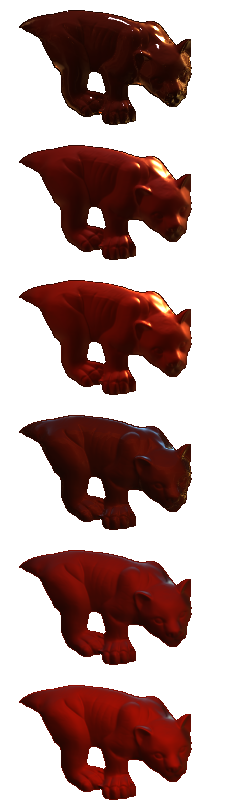}
\end{minipage}
\hfill
\begin{minipage}{.45\linewidth}
\includegraphics[height=\linewidth]{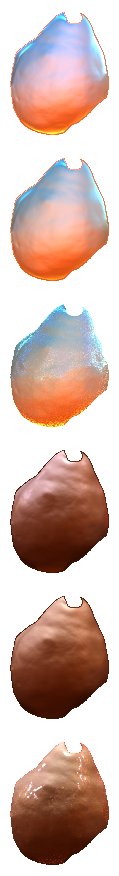}
\includegraphics[height=\linewidth]{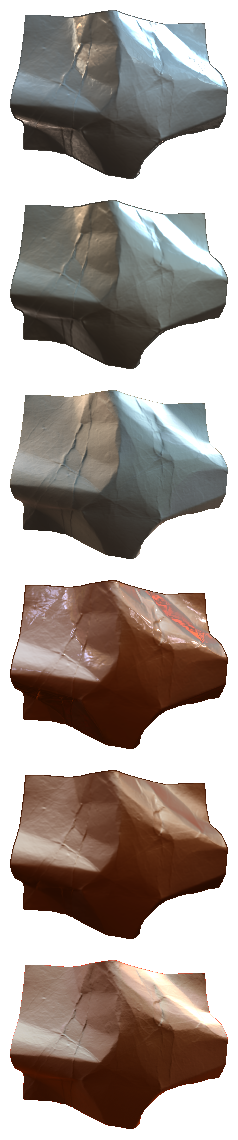}
\includegraphics[height=\linewidth]{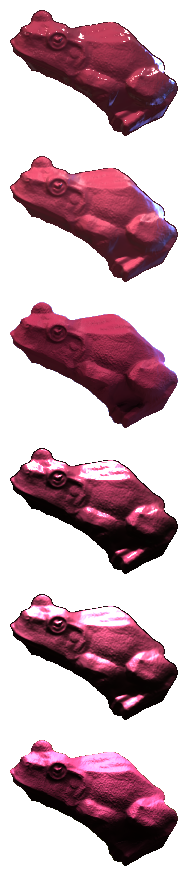}
\includegraphics[height=\linewidth]{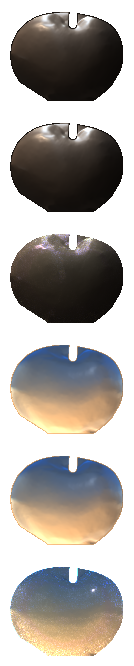}
\end{minipage}
\caption{Qualitative results on both versions of our dataset Materials are estimated using our blind optimized (opt) and optimized+regressed (reg) methods, and compared to ground truth (true). The true original illumination image is also the input for estimating material.  Notice that our technique can recover both glossy and matte materials, performs well even for these complex shapes. Our method attains visually pleasing results even for complex reflectance functions not encoded by our model (e.g. measured dataset) even in {\it new}  lighting conditions.
\vspace{-10mm}
}
\label{fig:qualitative}
\end{figure*}

\begin{figure*}
\begin{minipage}{.10\linewidth}
\hfill
\begin{sideways}
{ \hspace{2mm} \underline{Median} \hspace{3mm} \underline{True} \hspace{6mm} \underline{Best} \hspace{8mm} }\end{sideways}
\begin{sideways}
{ \hspace{1mm} opt \hspace{2mm} reg \hspace{14mm} reg \hspace{2mm} opt }
\end{sideways}
\end{minipage}
\begin{minipage}{.40\linewidth}
\centerline{\large\underline{Fit dataset materials}}
\includegraphics[width=\linewidth]{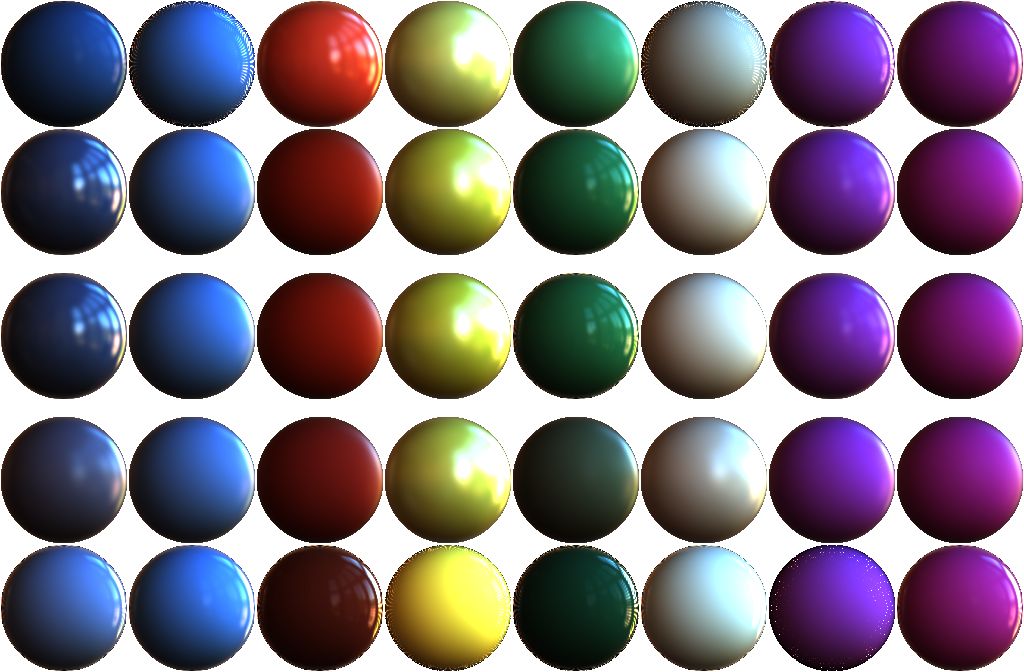}
\end{minipage}
\hfill
\begin{picture}(20,100)
\thicklines
\put(10,-60){\vector(0,1){50}}
\put(10,52){\vector(0,-1){50}}
\end{picture}
\hfill
\begin{minipage}{.40\linewidth}
\centerline{\large\underline{Measured dataset materials}}
\includegraphics[width=\linewidth]{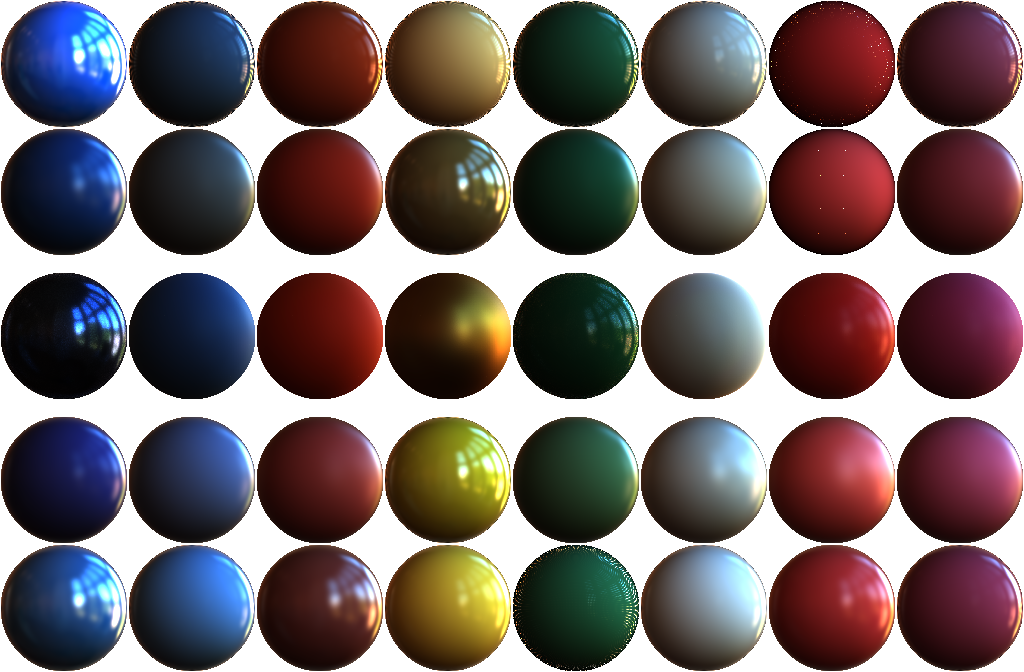}
\end{minipage}
\caption{Comparison of estimated materials rendered in novel lighting. The true materials lie on the middle row alongside our per-material best and median optimized (opt) and regressed (reg); arrows indicate the direction in which materials should improve. We achieve very good results for input images that are well described by our model in the fit dataset (rows 2 and 4 generally look like row 3), and even in many cases for measured BRDFs. However, low-order model bias prevents our method from capturing certain materials well (e.g. column 4; measured dataset).
}
\label{fig:spheres}
\end{figure*}

\boldhead{Results}
We generate results using the optimization procedure described in Sec~\ref{sec:est}, followed by our bias regression method as in Sec~\ref{sec:bias}. Bias prediction functions are learned through leave-one-out cross validation. 

In this section, we report results from our optimization technique ({\bf blind optimized}), and after bias regression ({\bf optimized+regressed}). For comparison, we compute a {\bf  baseline} material estimate which computes the $R_d$ by averaging the image pixels in each channel, and $R_s$ and $r$ from the average found in our material dataset, and then regress and apply bias predictors to the baseline estimates ({\bf baseline+regressed}). We also compare to materials achieved by our optimization assuming the shape and illumination are known\footnote{Known lighting is fit to our parameterization and may still be some distance from ground truth} and fixed ({\bf known shape+light}); hence only the material is optimized. Results using this procedure gauge the difficulty of our optimization problem, and shows how much our optimization can improve with more sophisticated initialization procedures.

On our fit dataset,  our full method (optimized + regressed) is capable of recovering highly accurate material parameters. Figure~\ref{fig:mat_err} plots the true material from our ``fit'' dataset against our estimated parameters for each of the 400 images. A perfect material estimate would lie along the diagonal (dashed line). Overall, we see a linear trend in our diffuse results, and that our bias regression can significantly improve our optimized estimates of specular reflectance and specular lobe size (and even better than shape+light).

We also develop two ways of measuring visual error in our materials. We define {\bf original illumination} as the average pixel error from comparing the input image with the image produced by rendering our {\it estimated} material onto the {\it true} shape in the {\it true} lighting (which are known for our all images in our dataset). This is a harsh test, as any errors in material must manifest themselves once rendered with the true shape and light. The second metric ({\bf cross rendered}) is even more telling: we compare renderings of the input object with the a) true material and b) our estimated material in six novel illumination environments not present in our dataset and compute average pixel error. This measure exposes material errors across unique, unseen illumination.

Using these measures, our full method achieves low error for both versions of our dataset. Figure~\ref{fig:rendererror} shows these error measures for three different metrics (per-pixel L2 and L1 norms, and absolute log difference), and optimized+regressed performs the best overall for both datasets.  This indicates that both our optimization and regression are crucial components, and one is not dominating inference since optimized+regressed consistently outperforms baseline+regressed. Known+shape light also performs well, indicating that our optimization procedure might improve if better initializations are available.

We demonstrate that in many cases our method can do very well at visually reproducing both measured and fit reflectances, even in novel illuminations. We show qualitative results for both versions of our dataset in Figs~\ref{fig:qualitative} and~\ref{fig:spheres} -- these are some of our best and median results.  Our material estimates are typically visually accurate in original and novel illumination, even for many of the measured BRDFs in our measured dataset. We also observe that our regression generally helps for both datasets, indicating that our learned bias predictors may generalize to complex materials and real-world images. However, it is clear that our results degrade for complex reflectance functions that lie well outside our model (Fig~\ref{fig:spheres}, measured dataset columns 1+4).

\begin{figure*}
\begin{minipage}{.34\linewidth}
  \begin{minipage}{.49\linewidth}
  \centerline{Original light}
  \centerline{\small \hspace{3mm} true \hspace{4.5mm} estimated}
  \includegraphics[width=\linewidth]{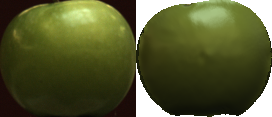}
  \centerline{ 
  \includegraphics[width=0.7\linewidth]{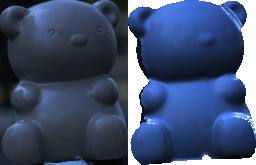}
  }
  \end{minipage}
  \begin{minipage}{.49\linewidth}
  \centerline{Novel light}
  \centerline{\small \hspace{3mm} true \hspace{4.5mm} estimated}
  \includegraphics[width=\linewidth]{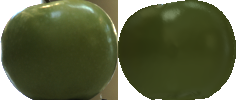}
  \centerline{ 
  \includegraphics[width=0.8\linewidth]{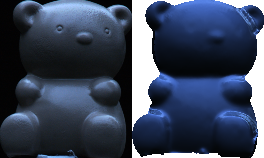} 
  }
  \end{minipage}
\end{minipage}
\hfill
\begin{minipage}{.25\linewidth}
  \begin{minipage}{.49\linewidth}
  \centerline{Original light}
  \centerline{\small \hspace{4mm} true \hspace{1mm} estimated}
  \includegraphics[width=\linewidth]{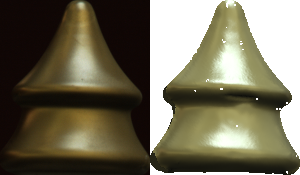}
 \centerline{ 
 \includegraphics[width=0.65\linewidth]{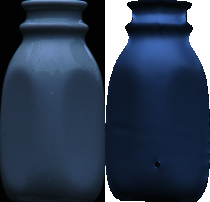}
 }
  \end{minipage}
  \begin{minipage}{.49\linewidth}
  \centerline{Novel light}
  \centerline{\small \hspace{4mm} true \hspace{1mm} estimated}
  \includegraphics[width=\linewidth]{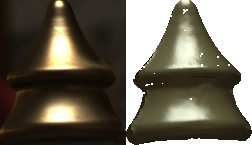}
  \centerline{
  \includegraphics[width=0.65\linewidth]{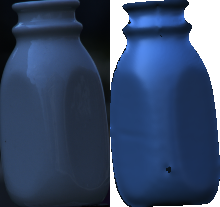}
  }
  \end{minipage}
\end{minipage}
\hfill
\begin{minipage}{.38\linewidth}
  \begin{minipage}{.49\linewidth}
  \centerline{Original light}
  \centerline{\small \hspace{2.5mm} true \hspace{4.5mm} estimated}
  \includegraphics[width=\linewidth]{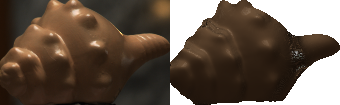}
  \includegraphics[width=\linewidth]{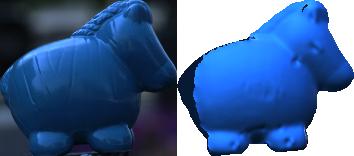}
  \end{minipage}
  \begin{minipage}{.49\linewidth}
  \centerline{Novel light}
  \centerline{\small \hspace{2.5mm} true \hspace{4.5mm} estimated}
  \includegraphics[width=\linewidth]{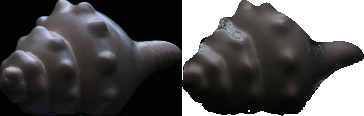}
  \includegraphics[width=\linewidth]{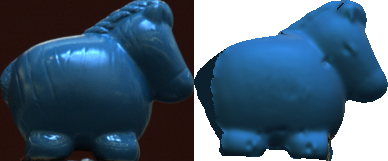}
  \end{minipage}
\end{minipage}
\caption{Results on real data from the Drexel Natural Illumination dataset. This dataset contains real images and corresponding ground truth shape and lighting information. We estimate materials from one picture, and render the material using the true shape and light for the original illumination and another illumination from the dataset (novel light); we compare to the real picture of the object in both scenes (original and novel). Even in the presence of slight spatial variation (e.g.  top left; apple) and complex reflectance (top middle) our method can still recover decent estimates. Still, addressing these issues is key to generalizing our method's applicability.
}
\label{fig:natgeom}
\end{figure*}

Finally, we demonstrate our method's capability on {\it real images} from the Drexel Natural Illumination dataset in Fig~\ref{fig:natgeom}. Our model appears somewhat robust to spatially varying reflectance in these images, but suffers from the complexity of the imaged reflectances and because we assume only a single material is present; this suggests ideas for future work.

\subsection{Inhomogeneous materials}
\label{sec:eval:sv}

For ground truth evaluation, we again use the {\bf measured dataset}. We use our mixture estimation procedure to estimate $k=\{2, 3\}$ materials\footnote{We use a spatial mixture for homogeneous materials as our mixture maps generalize current literature. They capture spatial variation in material (as in~\cite{Goldman:tpami:2010}), but we use them to also encode any kind of surface variation not well-captured due to long-standing SFS assumptions.} for each dataset image, and compare to the results in of our method for $k=1$. For additional comparison, we compute a baseline material estimate by clustering the image into $k$ components (using $k$-means); computing diffuse albedo (per component) by averaging the image pixels in each channel, and the specular components are fixed to a small yet reasonable value.

We measure error by rendering our {\it estimated} material onto the {\it true} shape in the {\it true} lighting (which are known for our all images in the dataset), and compare this to the input image. We do the same test, but for six novel lighting environments not found in the dataset (e.g. estimated material versus true material in novel light). We denote these as ``orig'' and ``cross'' lighting respectively. These are harsh tests of generalization, as any errors in material must manifest themselves once rendered with the true shape and light, and the ``cross'' measure exposes material errors across unique and unseen illumination.

\begin{figure}
\includegraphics[width=\linewidth]{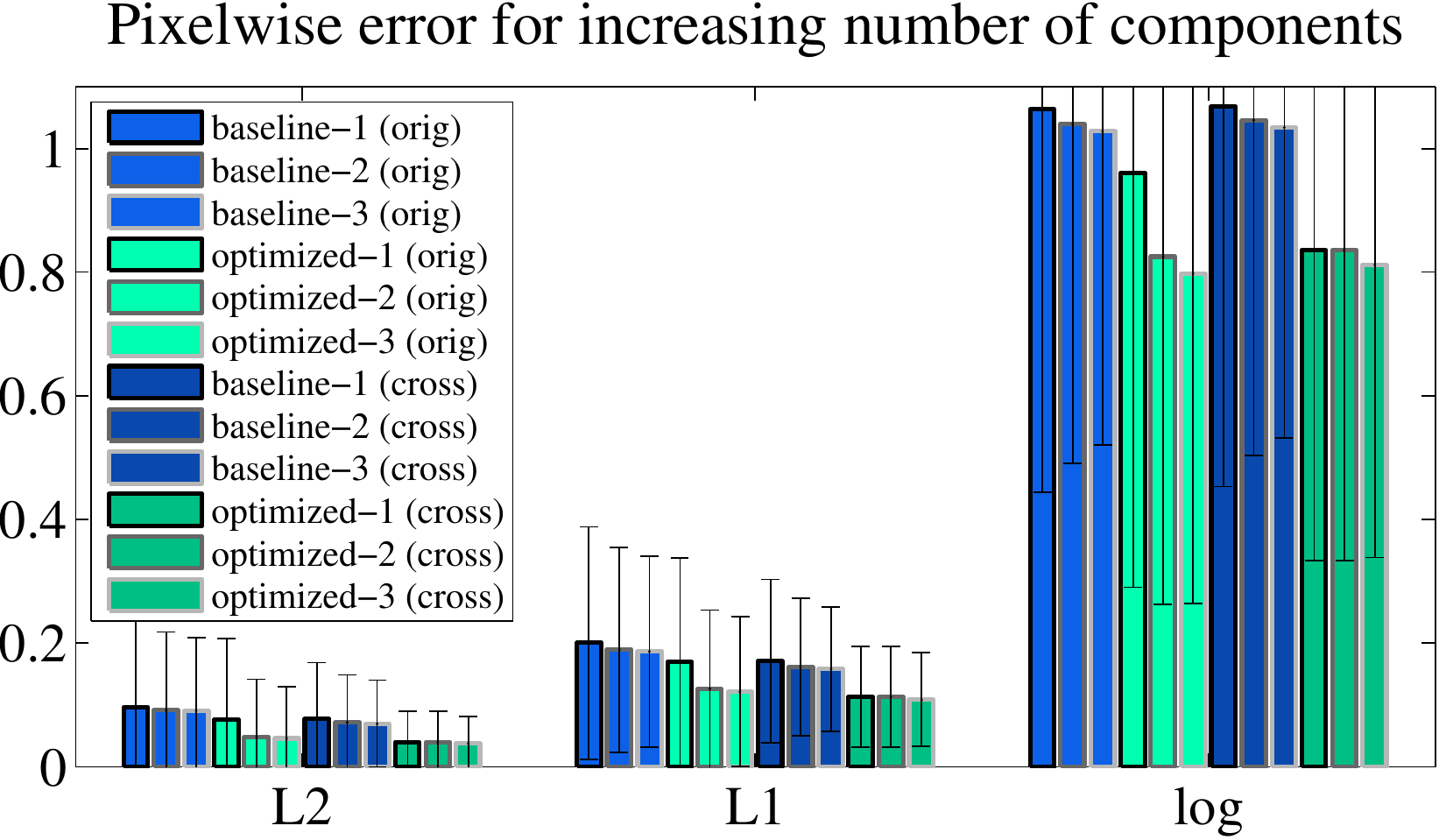}
\caption{Quantiative results on our ``measured'' dataset. 
Our mixture materials (optimized-$\{2,3\}$) consistently outperform single material estimation (optimized-$1$); see text for details.
}
\label{fig:quantPBRTdata}
\end{figure}

Fig~\ref{fig:quantPBRTdata} shows quantitative results averaged over the entire dataset for L2, L1, and absolute log difference error metrics. Our mixture materials (optimized-$\{2,3\}$) consistently outperform single material estimation (optimized-$1$), and are always better than the baseline estimates. 

\begin{figure*}[t]
\begin{minipage}{\linewidth}
  \hfill
  \begin{minipage}{.45\linewidth}
  \centerline{\Large Original light}
  \centerline{\hspace{0mm} 1 material \hspace{3mm} 2 materials \hspace{3mm} 3 materials \hspace{7.5mm} true \hspace{5mm}}
  \includegraphics[width=\linewidth]{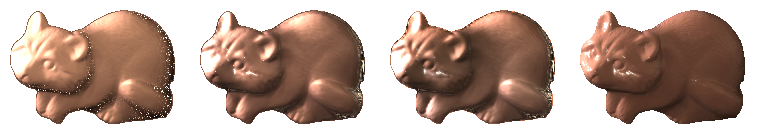}
  \includegraphics[width=\linewidth]{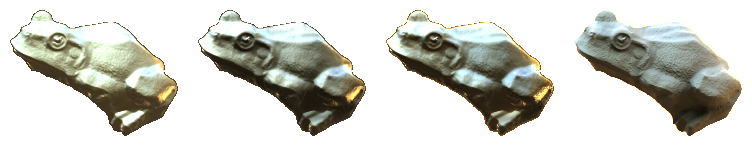}
  \end{minipage}
  \hfill
  \begin{minipage}{.45\linewidth}
  \centerline{\Large Novel light}
  \centerline{\hspace{0mm} 1 material \hspace{3mm} 2 materials \hspace{3mm} 3 materials \hspace{7.5mm} true \hspace{5mm}}
  \includegraphics[width=\linewidth]{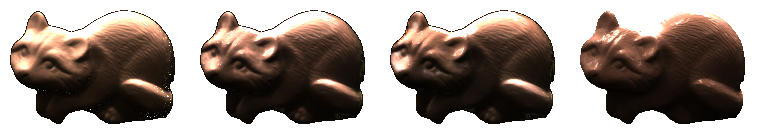}
  \includegraphics[width=\linewidth]{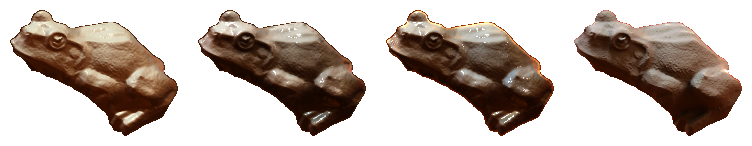}
  \end{minipage}
  \hspace*{\fill}
\end{minipage}
\caption{Best (top row) and median (bottom row) results from the ``measured dataset'' which contains physically rendered objects with measured BRDFs. Typically these materials are not well encoded by our low-order material model with 1 mixture component, but increasing the number of mixture components improves re-rendering error. We show our estimated materials for one, two, and three mixture components, and compare these to the ground truth result (also the input image) in both the original and novel illumination environments.}
\label{fig:qualPBRTdata}
\end{figure*}

We observe a similar trend in our qualitative results (Fig~\ref{fig:qualPBRTdata}). Because we are attempting to estimate true, measured BRDFs which may lie outside of our 5-parameter material model, estimation may not work well with a single material. However, by adding multiple materials, we typically get improved results, even in novel illumination. This indicates that our mixture weights are typically robust to shading artifacts such as shadows and specularities. It is clear that adding more components helps, although the distinction between $k=2,3$ is subtle (both qualitatively and quantitatively).

\section{Applications}
\label{sec:generate}

Once we have decomposed an image into its materials and spatial mixing weights, we can apply this intrinsic material information to new surfaces as in Fig~\ref{fig:teaser}. Applying the materials (microstructure) to a novel object is straightforward, but transferring the mixture weights (macrostructure) can be challenging in certain cases (e.g. when a mapping from one surface to another is not easily computed).

We propose a straightforward solution: choose a small patch of the image defined by the mixture weights that is nearly fronto-parallel (determined from our predicted surface normals; to avoid foreshortening), and synthesize a larger texture (seeded with the small patch) using existing methods; e.g.~\cite{Efros99}.  Then, map the surface of the object that the material will be transferred to onto a plane (also using existing methods; e.g.~\cite{Sheffer:2006}); this mapping defines correspondences between the synthesized mixture weights and the new mesh. We generate all of our transfer/generation results using this technique, and more sophisticated methods are clear directions for future work.

We also propose a generative material modeling strategy: besides transferring a complete mixture material, we can combine estimates from multiple images to create new materials (e.g. materials from one and mixture weights from another, and so on).

\begin{figure*}[t]
\begin{center}
\begin{minipage}{.16\linewidth} \hfill \end{minipage}
\begin{minipage}{.03\linewidth} \hfill \end{minipage}
\begin{minipage}{.64\linewidth}
  \begin{center}  \hspace{0.09\linewidth}
  glass \hfill base \hfill  texture \hfill fiber \hfill gloss \hspace*{0.075\linewidth} \\
  \hspace{0.07\linewidth}
  \includegraphics[width=0.1\linewidth]{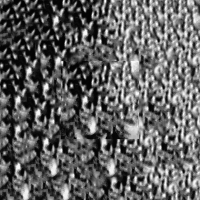} \hfill
  \includegraphics[width=0.1\linewidth]{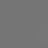} \hfill
  \includegraphics[width=0.1\linewidth]{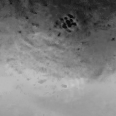} \hfill
  \includegraphics[width=0.1\linewidth]{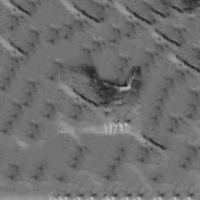} 
  \hspace*{0.07\linewidth}
  \includegraphics[width=0.1\linewidth]{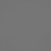} \hspace*{0.05\linewidth}
  \end{center}
\end{minipage} \\
\vspace{2mm}
\begin{minipage}{.16\linewidth}
       \begin{overpic}[width=0.59\linewidth]{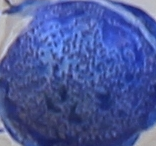}
      \put(1,3){ \color{white}blue-glass}
      \end{overpic} \\
       \begin{overpic}[width=0.59\linewidth]{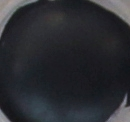}
      \put(4,3){ \color{white}gray-base}
      \end{overpic} \\
       \begin{overpic}[width=0.59\linewidth]{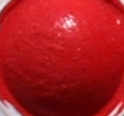} 
      \put(3,3){ \color{white}red-texture}
      \end{overpic} \\
 \end{minipage}
\begin{minipage}{.03\linewidth}  
    \hfill
  \begin{sideways}
      \hspace{3mm} red \hspace{11mm} gray \hspace{10mm} blue
  \end{sideways}
 \hspace*{2mm}
\end{minipage}
\begin{minipage}{.64\linewidth}
\vspace{-4mm}
  \includegraphics[width=\linewidth,clip=true, trim=0 480 0 0]{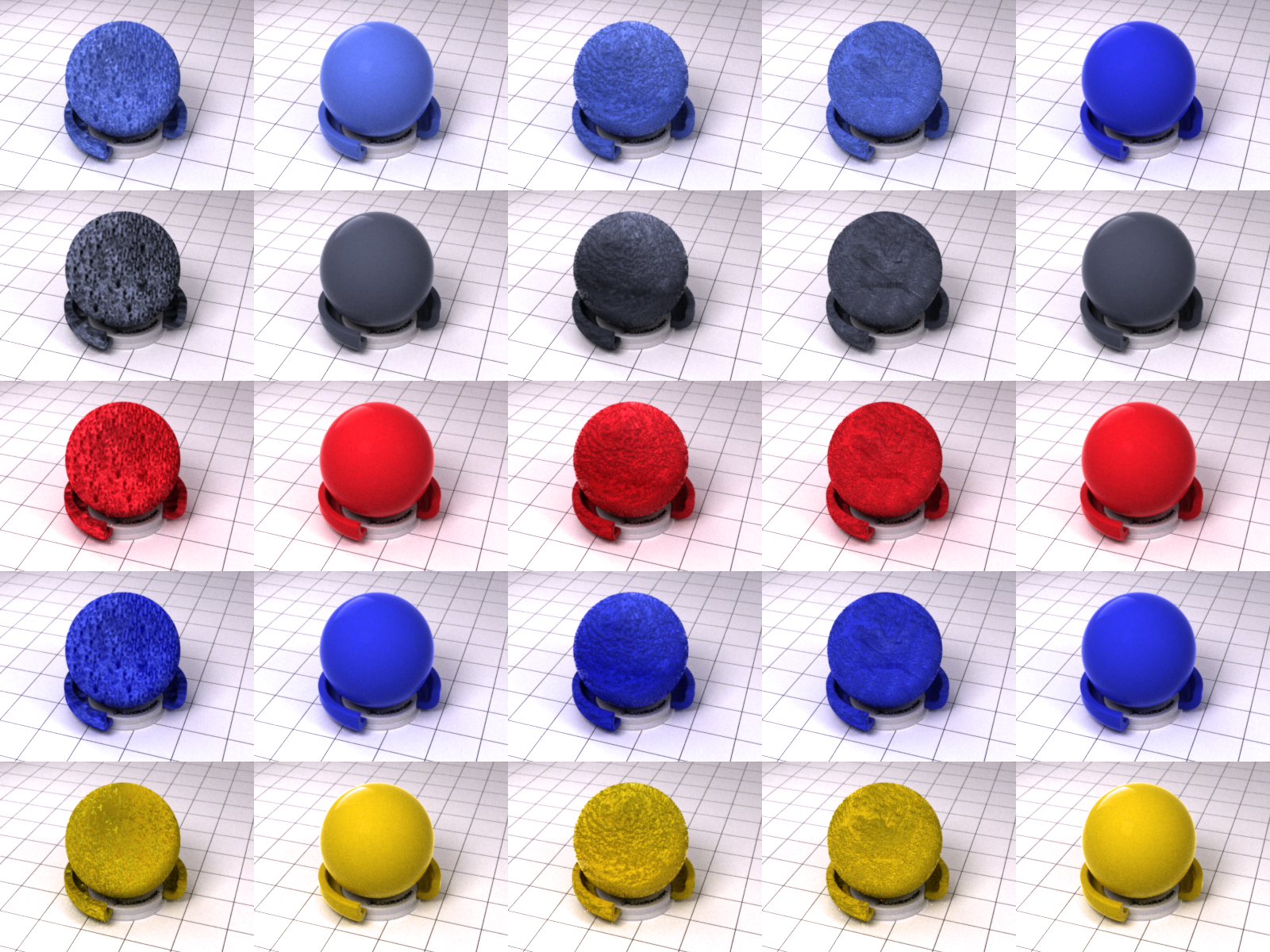}
\end{minipage}
\end{center}
\vspace{-3mm}
\caption{Material transfer and generation for several material "swatches" (hemispheres painted with different colors/mediums/coats). We decompose {\it single images} (on left) into two material components and a spatial mixture map. Then, we synthesize new materials by taking all combinations of the inferred materials and the derived mixture weights, and render these combinations onto spheres in novel illumination (using LuxRender: \urlwofont{http://luxrender.net}). Images along the diagonal show a transfer material result for a given picture on the left. The off-diagonals show the generative capabilities of our algorithm: by combining multiple decompositions (materials + mixing weights), we can generate new, unseen materials. We expect that full 3D textures will give better results, but it is currently impossible to estimate 3D textures from a single picture. Best viewed in color at high resolution.
}
\label{fig:renderStealGrid}
\end{figure*}

Generative results (as well as direct transfer results) are shown in Fig~\ref{fig:renderStealGrid}. We have decomposed four swatches from our dataset (all unique colors and mediums and spanning the three illumination environments in our dataset) using $k=2$ mixture components. We apply each set of materials to each synthesized mixture, and render the result onto spheres.  We assert that our estimated materials correspond to microstructure and mixing weights correspond to macrostructure, which appears correct for these results (microstructure varies vertically, macrostructure horizontally).

\section{Conclusion}

We have demonstrated a new technique for estimating spatially varying parametric materials from an image of a single object of unknown shape in unknown illumination, going beyond the typical Lambertian assumptions made by existing shape-from-shading techniques. Strong priors and low-order parameterizations of lighting and material are key in providing enough constraints to make this inference tractable. Such rigid parameterizations often lead to estimation bias, and we also present a simple yet powerful technique for removing this bias. 

Our results suggest that material recovery is not necessarily dependent upon the joint recovery of accurate shape and illumination; {\it as long as the shape and illumination are consistent with each other, materials can still be robustly estimated}. This is encouraging from a material inference standpoint, as even the best shape-from-shading algorithms still produce flawed estimates in many scenarios.

As far as we know, our method is the first to estimate parametric material models without assuming shape or illumination is known a priori. We believe that our method provides good initial evidence that solving this problem is in fact feasible, and provides a foundation for estimating materials from photographs alone.


Our decompositions can be transferred to new shapes, imbuing them with similar appearance as the input image. Furthermore, our decompositions are also generative, and can be used to create new materials by simultaneously transferring decompositions from multiple objects (e.g. mixing weights from one, materials from another). Our re-rendering results do not incorporate any information from our estimated surface normals, and the spatial frequency of our mixture weights are defined by the input image resolution (some artifacts visible in Fig~\ref{fig:teaser}); intelligently incorporating and up sampling these estimates are reasonable directions for future work.

\bibliographystyle{acmsiggraph}
\bibliography{materials,mix-materials}   

\end{document}